\documentclass[12pt]{iopart}
\usepackage{graphicx}
\usepackage{dcolumn}
\usepackage{bm}
\usepackage{hyperref}

\usepackage{color}
\definecolor{grey}{gray}{0.5}

\begin{document}

\title{Phase Space Tomography of Matter-Wave Diffraction in the Talbot Regime}

\author{S. K. Lee and M. S. Kim}
\address{QOLS, Blackett Laboratory, Imperial College London,SW7 2BW, UK}
\author{C. Szewc and H. Ulbricht}
\address{Physics and Astronomy, University of Southampton, SO17 1BJ, UK}
\ead{h.ulbricht@soton.ac.uk}

\begin{abstract}
We report on the theoretical investigation of Wigner distribution function (WDF) reconstruction of the motional quantum state of large molecules in de Broglie interference. De Broglie interference of fullerenes and as the like already proves the wavelike behaviour of these heavy particles, while we aim to extract more quantitative information about the superposition quantum state in motion. We simulate the reconstruction of the WDF numerically based on an analytic probability distribution and investigate its properties by variation of parameters, which are relevant for the experiment. Even though the WDF described in the near-field experiment cannot be reconstructed completely, we observe negativity even in the partially reconstructed WDF. We further consider incoherent factors to simulate the experimental situation such as a finite number of slits, collimation, and particle-slit van der Waals interaction. From this we find experimental conditions to reconstruct the WDF from Talbot interference fringes in molecule Talbot-Lau interferometry.

\end{abstract}

\maketitle

\section{Introduction}
The reconstruction of a quantum state in general is important to test potential quantum systems on their properties such as entanglement, superposition and coherence and on their applicability for emerging quantum technologies. Furthermore, massive and spatially extended quantum systems such as complex molecules and nanoparticles in quantum superposition are the proposed test embodiments for universal boundaries of the validity of quantum theory with strong relevance for future nanotechnology~\cite{nimmrichter2011testing, romero2011optically}. The wave nature of molecular motional states (as massive as ten C$_{60}$ molecules) has been demonstrated~\cite{Gerlich2011quan} and to make a step further, we now want to characterize the quantum state of motion of molecules in Talbot-Lau interferometry.

The wave function cannot be observed directly, but by using the Wigner distribution function (WDF) we have an alternative perspective on quantum dynamics as WDF is equivalent to the density matrix~\cite{wigner1932quantum,schleich2001}. Quantum states have the unique property that they \emph{can} generate negative values of this quasi-probability function. On the first hand the negativity of the Wigner function is seen as a proof of the quantum nature of the state under consideration. A fully reconstructed Wigner function contains the complete information about the measured state. The process to evaluate the state is called phase space tomography.

\begin{figure}
\centering
\includegraphics[scale=0.25]{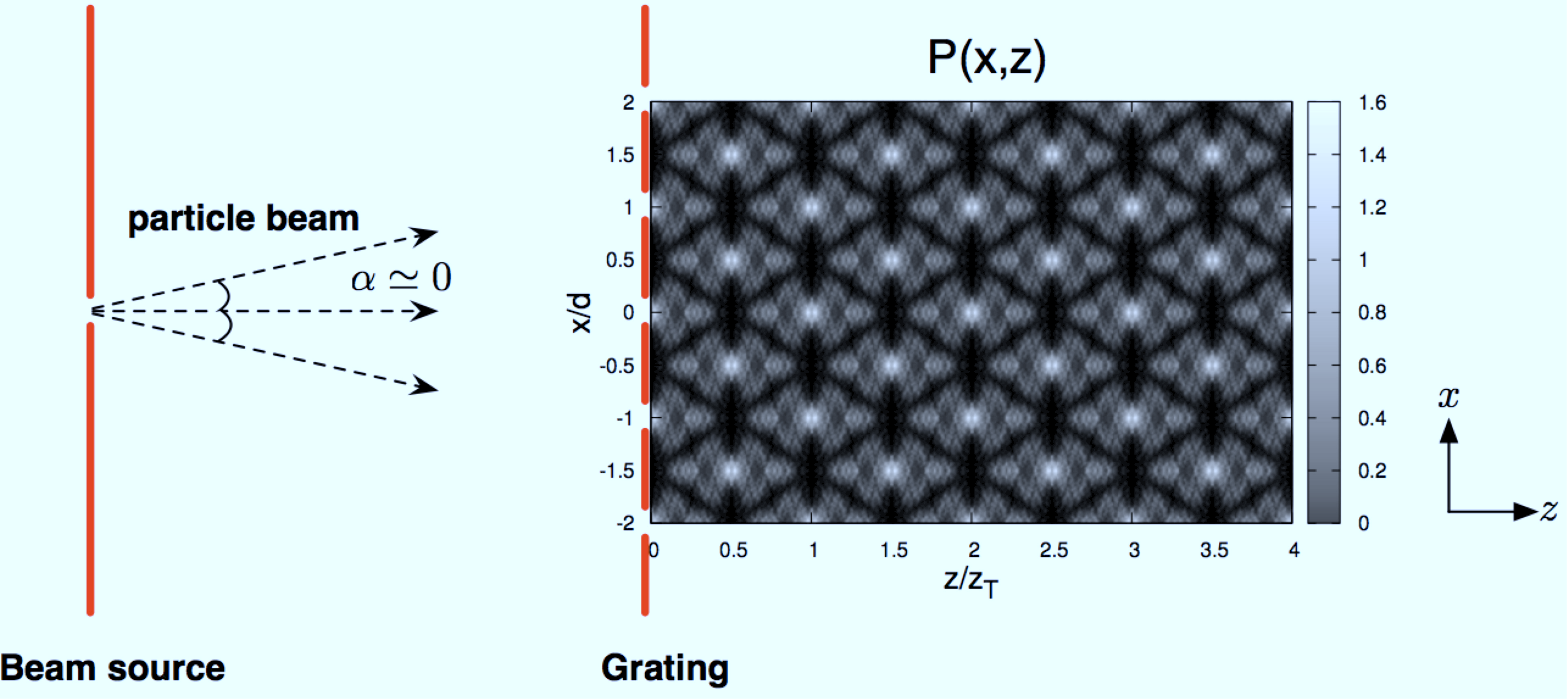}
\caption{\label{fig:talbot}The setup as used for simulations. The probability distribution for 2 Talbot length $z_T$, the quantum carpet, simulated with finite grating and grating opening fraction (slit width / period) of $f_o=0.3$ is illustrated. The reappearance of the grating self-images is the Talbot effect, which occurs if a periodic structure is coherently illuminated.}
\end{figure}

Phase-space tomography for the Wigner function was pointed out in
a general context by Bertrand and Bertrand~\cite{bertrand1987tomographic}and independently by Vogel and Risken~\cite{Vogel1989}, applied in photonics~\cite{leonhardt1997measuring} where quantum state tomography is an established experimental tool to quantify the quantum state of light~\cite{lundeen2008tomography}. It has been recently used to demonstrate the state squeezing of atomic Bose condensates~\cite{schmied2011tomographic}), and to quantify the motional quantum state of a trapped Be$^+$ ion~\cite{leibfried1996experimental}. An early experiment to characterize the quantum nature of atomic motion in de Broglie interference was the Wigner function reconstruction of meta-stable He atoms diffracted at a double slit~\cite{kurtsiefer1997measurement}, as theoretically proposed earlier~\cite{janicke1995tomography}. The Wigner function reconstruction has been also proposed to be of use to prove the quantum nature of the superposition of very massive particles~\cite{romero2011} and even macroscopic opto-mechanical systems~\cite{Vanner2011}. Furthermore, to prove entanglement in superconducting qubits, quantum state tomography has been applied~\cite{steffen2006measurement}.
In more technical terms the Wigner function is a quasi-probability distribution of states in phase space. In the case of de Broglie interference of molecules the quantum state is projected on to the spatial coordinate, which is the spatial number distribution of particles after the diffraction grating (see Fig.\ref{fig:talbot}). This spatial distribution is needed for different rotational angles of the quantum state in phase space. This rotation in phase space comes natural from free space propagation of a particle beam after diffraction and can be evaluated from measuring the spatial distribution at different distances after the grating. The inverse Radon transformation of collected data of density distributions will give the Wigner function. In the experiment the spatial coherence is prepared by another grating placed in front of the diffraction grating. Then many coherent sources constructively contribute to the same interference pattern due to the Lau effect. The multi grating configuration is the so-called \textit{Talbot-Lau} interferometer (TLI) and is explained in detail elsewhere~\cite{nimmrichter2008theory,brezger2002}.

Here, we discuss the Wigner function reconstruction of matter-waves in the near-field Talbot regime with illumination of the grating by a single coherent source. If the spatial coherence of the matter wave is high, which means that plane waves are reaching the grating, then the here presented Talbot simulations are valid for the Talbot-Lau scheme and reconstruction gives the same Wigner function.

We theoretically perform phase-space tomography of the center of mass motion quantum state of massive molecules in a near-field TLI~\cite{clauser1992}. Near-field means that the interference pattern is on the same size scale as the diffraction grating period. In contrast, the far-field (Fraunhofer) interference pattern are much larger than the grating (for illustration see Hornberger et al.~\cite{Hornberger2012}). Earlier investigations of de Broglie quantum states have dealt with far-field pattern reconstruction~\cite{kurtsiefer1997measurement,janicke1995tomography}. From light and matter-wave optics it is known that complex diffraction pattern can be expected in this near-field regime. Those structures are sometimes called quantum carpets~\cite{friesch2000,berry2001}. Recently optical quantum carpets have been experimentally observed by Case et al.~\cite{case2009realization}. Talbot carpets have been applied for computation as number factorization~\cite{clauser2008,schleich2008,gilowski2008gauss}.

\section{Theoretical Model}

\subsection{Phase Space Tomography}
The WDF $W(x,p)$ of a complex signal $\psi(x)$ is defined by \cite{wigner1932quantum}:
\begin{equation}
W(x,p)=\frac{1}{\pi}\int_{-\infty}^{\infty} \psi^*(x-x')\psi(x+x')e^{-i2 p x'}dx' ,
\label{eq:wdf}
\end{equation}
with momentum $p$ and position $x$. Throughout the paper we set $\hbar$=1. Experimentally, we measure the spatial intensity pattern which corresponds to the projection of the WDF onto the space coordinate. This is formulated by integration of the WDF over the momentum variable $p$:
\begin{equation}
P(x)=\int_{-\infty}^{\infty}  W(x,p) dp.
\label{eq:P}
\end{equation}
If the WDF is rotated by angle $\theta$ in phase space, it becomes
\begin{equation}
W_{\theta}(x,p)=W(x\cos\theta-p\sin\theta, x\sin\theta+p\cos\theta),
\end{equation}
and in analogy to Eq.~(\ref{eq:P}) the spatial intensity pattern $P_{\theta}(x)$ or the marginal probability can be obtained by
\begin{equation}
P_{\theta} (x)=\int_{-\infty}^{\infty}  W_{\theta}(x,p) dp .
\label{eq:Ptheta}
\end{equation}
This intensity pattern with various angles of rotation can be obtained from diffraction pattern like a quantum carpet for a grating with an infinite number of slits. Phase space tomography is based on the transformation of Eq.~(\ref{eq:Ptheta}) resulting in a reconstructed WDF:
\begin{equation}
W(x,p)=\frac{1}{4\pi^2}\int_{-\infty}^{\infty}  dx'  \int_{0}^{\pi} d\theta P_{\theta}(x') \int_{-\infty}^{\infty}   dr|r| e^{ir(x'-x\cos\theta-p \sin\theta)},
\label{eq:radon}
\end{equation}
which is the \emph{inverse Radon transformation}. Eq.~(\ref{eq:Ptheta}) can be written as
\begin{equation}
P_{\theta}(x)=\sum_n p_n|\psi_{n,\theta}(x)|^2,
\end{equation}
with
\begin{equation}
\psi_{n,\theta}(x)=\frac{1}{\sqrt{2\pi i \sin\theta}} \int_{-\infty}^{\infty} dx' e^{-i(\frac{x}{\sin\theta}x' -\frac{1}{2}\cot\theta x'^{2})}\psi_n(x'),
\label{eq:fresnelwave}
\end{equation}
which represents the fact that the active rotation of the Wigner function can be done by fractional Fourier transformation in Fresnel diffraction theory and was derived in~\cite{janicke1995tomography}.

\subsection{Wigner Function Reconstruction for Free Space Propagation}

A full reconstruction of the WDF requires spatial probability distributions $P_{\theta}(x)$ for every angle $\theta$ between 0 and $\pi$. The easiest way to rotate the WDF is the free space propagation of the particles. The rotation angle then depends on the distance $z$ between the diffraction grating and the detector. The free propagation of the diffracted wave function is according to Fresnel diffraction theory given by~\cite{Hornberger2012}:
\begin{equation}
\psi(x,z)=\frac{1}{\sqrt{i\lambda z }}\int_{-\infty}^{\infty} dx' e^{i\frac{k}{2z}(x-x')^2}\psi(x',0),
\label{eq:fdt}
\end{equation}
where $k=2\pi/\lambda$ is equivalent to momentum,with $\hbar=1$, and $\lambda$ is the de Broglie wavelength. We rescale the $x$-axis and get the expression:
\begin{equation}
\psi(\frac{x}{s},z)=e^{-ik\frac{x^2}{2zs^2}} \int_{-\infty}^{\infty} dx' e^{-ik(\frac{xx'}{sz}-\frac{x'^2}{2z})}\psi(x',0).
\label{eq:fresnelevolution}
\end{equation}
Comparing Eqs. (\ref{eq:fresnelwave}) and (\ref{eq:fresnelevolution}) gives $s=k\sin(\theta)/z$, for the rescaling of the $x$-axis, and $\cot(\theta)=k/z$ for the dependency of the rotation angle $\theta$ from the distance $z$. As a first result we find that free space propagation does not lead to a full rotation over $\pi$ for finite $z$. Sufficient rotation to fully reconstruct the WDF can be achieved by a lens. Such a lens has been implemented for atomic matter-waves by a Fresnel zone plate~\cite{carnal1991imaging, reisinger2009poisson} or a standing light wave~\cite{sleator1992imaging}. In~\cite{janicke1995tomography} it was shown theoretically that by using a lens a $\pi$/2-rotation, and therefore the Fourier transform is accessible for finite values of $z$. Since the realization of such a lens has not been demonstrated for molecular matter-waves, we here consider the simplest case of rotation, which is free space propagation. However it is possibility to increase the accessible angle of rotation and therefore the amount of information about the quantum state by additional symmetry assumptions on the investigated WDF~\cite{pfau1997partial}, which we will use in Sec. (\ref{rotationangle}). We will investigate the \emph{partial} reconstruction of WDF in the Talbot regime for $\Theta$ between $0$ and $\pi/2$. As a figure of merit we use the appearance of negativity of WDF.

Technically, we numerically reconstructed WDF using the filtered back-projection algorithm~\cite{Herman80} for the inverse Radon transformation:
\begin{equation}
W(x,p)=\int_{0}^{\pi}d\theta \int_{-x_m}^{x_m} dx' P_\theta(x') g(x'-x\cos\theta -p\sin\theta),
\end{equation}
with
\begin{equation}\label{eq:cutoff}
g(x)\approx 2(-\frac{1}{x^2}+\frac{\cos(r_c x)}{x^2}+\frac{r_c\sin(r_c x)}{x}),
\end{equation}
which is approximated from $g(x)=\int_{-\infty}^{\infty} dr |r| e^{irx}$, where $x_m$ is a real range of the transverse $x$-axis and $r_c$ is a cut-off frequency.

\subsection{WDF of Ideal Quantum Carpet and Talbot Effect}
The WDF for a quantum superposition state as in the classic double slit arrangement, a cat state, is well known. Here, we are interested in the WDF of an ideal quantum carpet, the near-field wave diffraction pattern after a grating with many slits. We start with a grating with an infinite number of slits, as it gives a nice analytical expression. The wave function $\psi(x)$ after the grating is:
\begin{equation}
\psi(x)= t_c(x)\otimes \sum_{n=-\infty}^{\infty} \delta(x-nd),
\label{eq:combf}
\end{equation}
where $d$ is the grating period, $t_c(x)$ is the grating transmission function for a single slit ($-d/2\leq x\leq d/2$), and $\otimes$ denotes convolution: $(f\otimes g)(x)=\int_{-\infty}^{\infty} f(x-x')g(x')dx'$. An infinite train of delta function, a comb function, can be expressed as  $\frac{1}{d} \sum_{n=-\infty}^{\infty}  e^{i2\pi n \frac{x}{d}} $. With that we rewrite Eq.~(\ref{eq:combf}) to be:
\begin{equation}
\psi(x)=\sum_{n=-\infty}^{\infty}A_n e^{i2\pi n \frac{x}{d}},
\end{equation}
where $A_n=d^{-1}\int_{-d/2}^{d/2} dx \,t_c(x) e^{-i2\pi n x/d}$.
The wave function along $z$ is then in form of Eq.(\ref{eq:fresnelevolution}) given by:
\begin{equation}
\psi(x,z)\sim \sum_{n=-\infty}^{\infty} A_n e^{-i2\pi n\frac{x}{d}} e^{-i\pi n^2 \frac{z \lambda}{d^2}}.
\end{equation}
The second exponential term generates a periodic appearance of the identical pattern at different positions in $z$-direction. This is the Talbot effect. Self-images of the grating revive at multiple of the Talbot distance $z_T = 2\frac{d^2}{\lambda}$. The probability distribution of this Talbot effect is shown in Fig.\ref{fig:talbot}. We note that self images are also revive at multiple of $z_T/2$ with a $d/2$ displacement in x-direction. It is this $d/2$ shift which carries non-classical information about the motional state and generates negative values of WDF. WDF reconstruction will therefore need to include especially such $z$-positions. In the following we define the diffracted wave function in units of the Talbot distance, since the self-image is repeated at multiple of $z_T$. We note that the number of near-field Talbot self-images typically depends on the number of contributing (coherently illuminated) grating slits. Experimentally, five Talbot lengths $z_T$ are easy to achieve~\cite{Hornberger2012}.

\begin{figure}[t]
(a)\hspace{7cm}(b)\\
\includegraphics[scale=0.65]{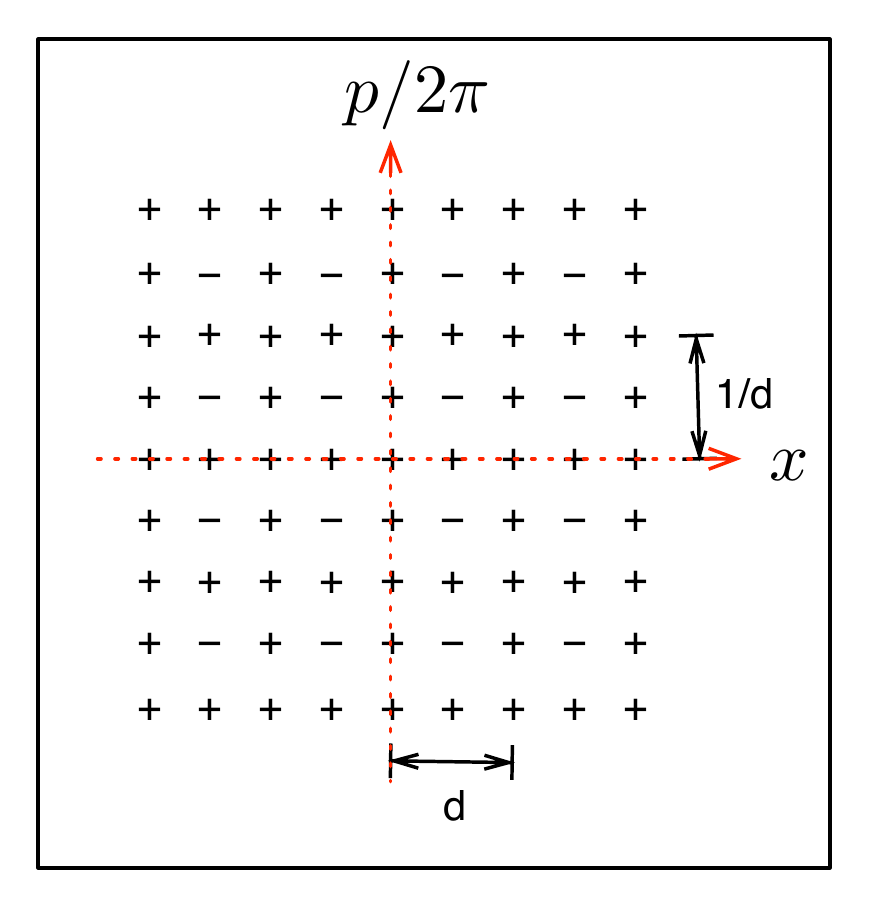}
\includegraphics[scale=0.70]{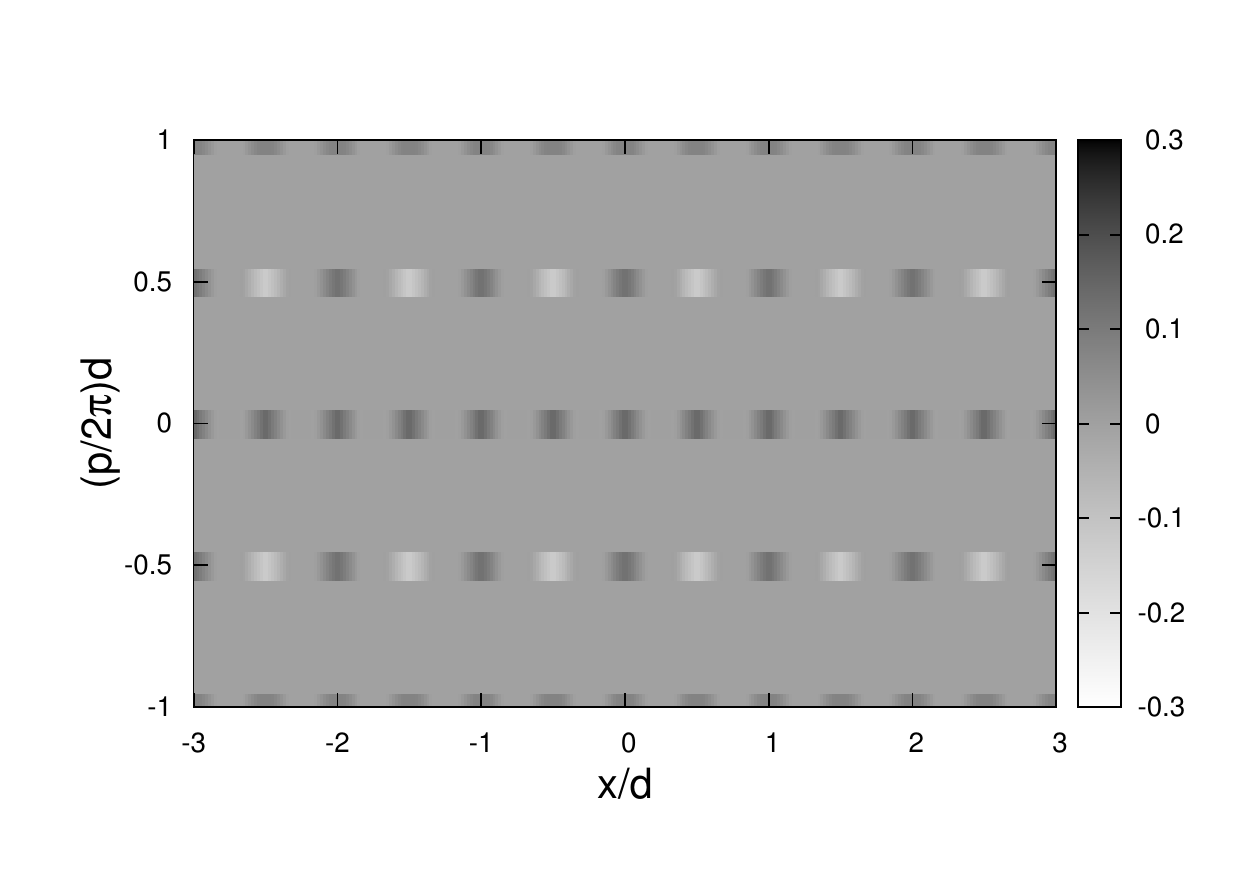}
\caption{Exact Wigner Distribution Function of (a) $\delta$-function comb wave and (b) rectangular wave. Those correspond to the WDF of infinite gratings. The symbols +, - in (a) indicate positive, negative delta peaks and the bar in (b) indicates gray scaled of WDF in the contour plot.}
\label{fig:wdfcomb}
\end{figure}

The exact WDF for the infinitely periodic input is obtained from Eq.~(\ref{eq:wdf}) as
\begin{equation}
W(x,p)=\sum_{n=-\infty}^{\infty}\sum_{n'=-\infty}^{\infty}A_nA^\ast_{n'} e^{i2\pi (n-n') \frac{x}{d}} \delta(p/2\pi-\frac{n+n'}{2d}).
\label{eq:WDFexact}
\end{equation}
The WDF of the $\delta$-comb function, which is the mathematically simplest guess, is obtained by substituting $m=n+n'$, and setting $A_n=1/d$:
\begin{eqnarray}
W(x,p)&=&\frac{1}{d^2} \sum_{n=-\infty}^{\infty} e^{i2\pi (2n) \frac{x}{d}}\sum_{m=-\infty}^{\infty}e^{-i2\pi (m) \frac{x}{d}}  \delta(p/2\pi-\frac{m}{2d}) \nonumber \\
&=& \frac{1}{2d} \sum_{n=-\infty}^{\infty} \sum_{m=-\infty}^{\infty} (-1)^{nm} \delta(x-\frac{nd}{2})  \delta(p/2\pi-\frac{m}{2d}),
\label{eq:WDFcomb}
\end{eqnarray}
and plotted in phase space as shown in Fig.~\ref{fig:wdfcomb}(a). The WDF is the sum of $\delta$-function in momentum, as seen in Eq.~(\ref{eq:WDFexact}). This delta function peaks, however, do not appear in experimental physical system where we have a finite number of slits of finite width. A grating with finite slit width is formulated with the Fourier coefficient $A_n=f_o sinc(n\pi f_o)$ and is plotted in Fig.~\ref{fig:wdfcomb}(b). We note the negative peaks (bright spots) in the WDF, showing the non-classical character of the state.  We observe that the peak width in momentum becomes narrower as the number of slits increases.

\section{Simulations: Limits for Finding Negativity in WDF}
Here we simulate the effects of experimental limitations on the quality of the reconstructed WDF, which include:
\begin{itemize}
\item Range of rotation angles,
\item Spatial detector resolution,
\item Incoherent source (in terms of collimation of molecule beam, spatial coherence),
\item Finite gratings,
\item Van der Waals interaction (interaction between molecules and material gratings),
\item Visibility.
\end{itemize}
We will conclude each section with the feasibility of these limits for the experiment. For numerical simulations, the grating period $d$ is set as unit length and $1/d$ is set as unit momentum.  All parameter ($x,z,\lambda$, $p$) become dimensionless by rescaling: $x \rightarrow x/d$, $z\rightarrow z/d$, $\lambda\rightarrow\lambda/d$, and $p\rightarrow pd$. The wavelength is chosen to be $\lambda=10^{-5}d$ and the open fraction of the gratings is set to $f_0$=0.3. This choice of parameter adapts the simulation to molecule interferometry experiments, which are in the centre of our interest~\cite{brezger2002}. However our results are universal and can easily be tuned to represent the same diffraction effects in the Talbot regime for other electromagnetic or matter waves. The cut off frequency $r_c$ as defined in Eq.~(\ref{eq:cutoff}) was optimized to $r_c=30$ to show the structure of WDF without high frequency computational noise. The WDF is reconstructed basing on the normalized probability distribution and will be plotted as contour maps in all following figures where the gray scale bar indicates the value of WDF. The scale bar allow for comparison of the different effects as the intensity scale has been normalized for each effect. For some WDF plots we show a cross section of the contour plot to visualize the negativity.

\subsection{Rotation Angle}\label{rotationangle}
\begin{figure}
(a)\hspace{8cm}(b)\\
\includegraphics[scale=0.65]{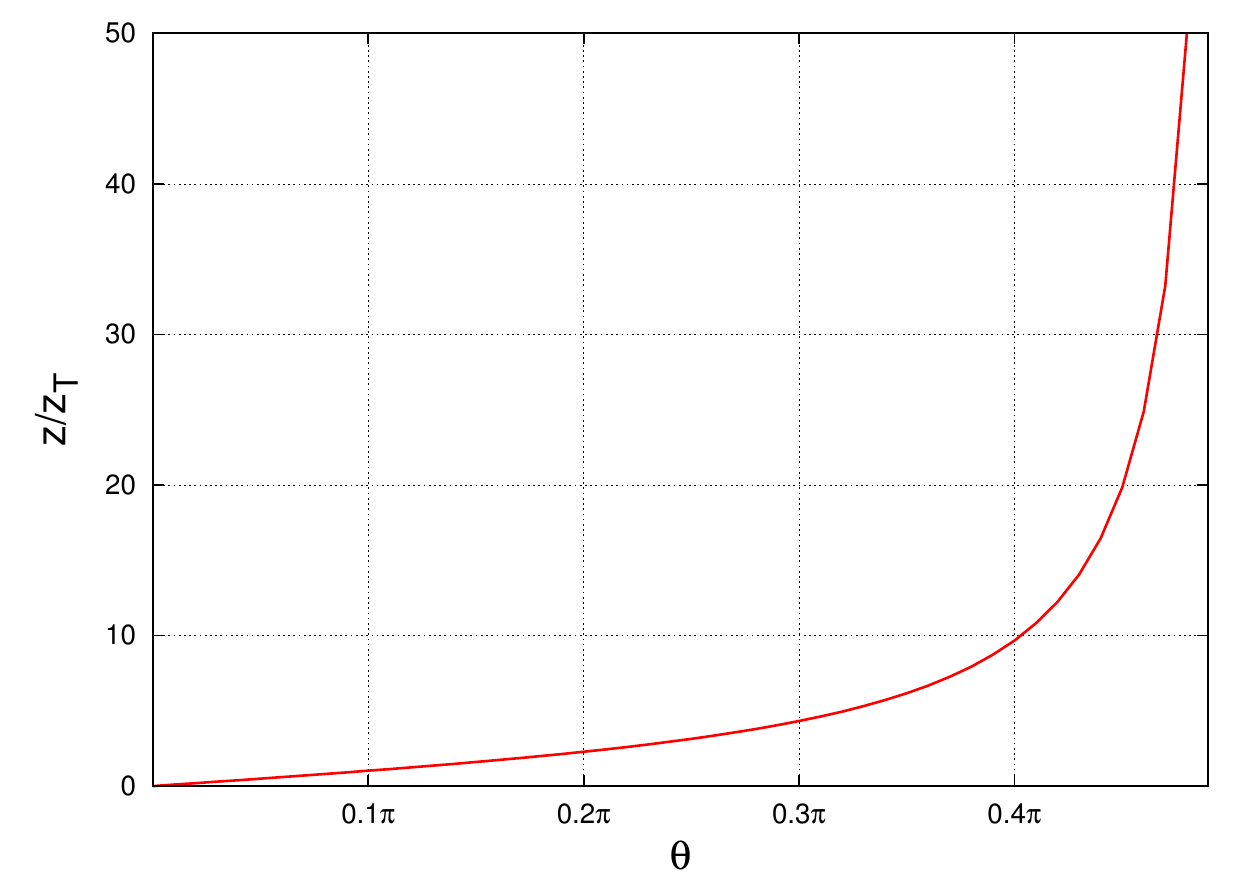}
\includegraphics[scale=0.65]{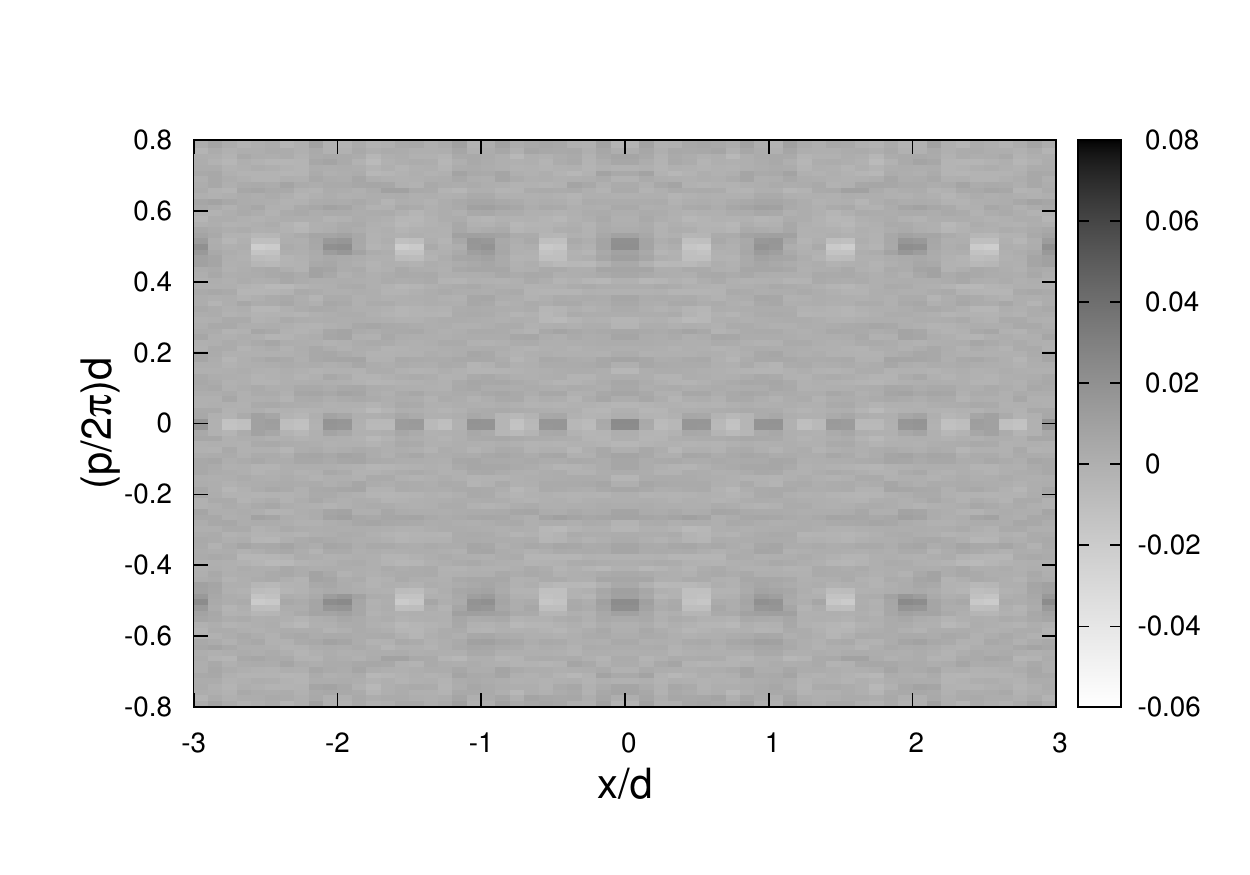} \\
(c)\hspace{8cm}(d)\\
\includegraphics[scale=0.65]{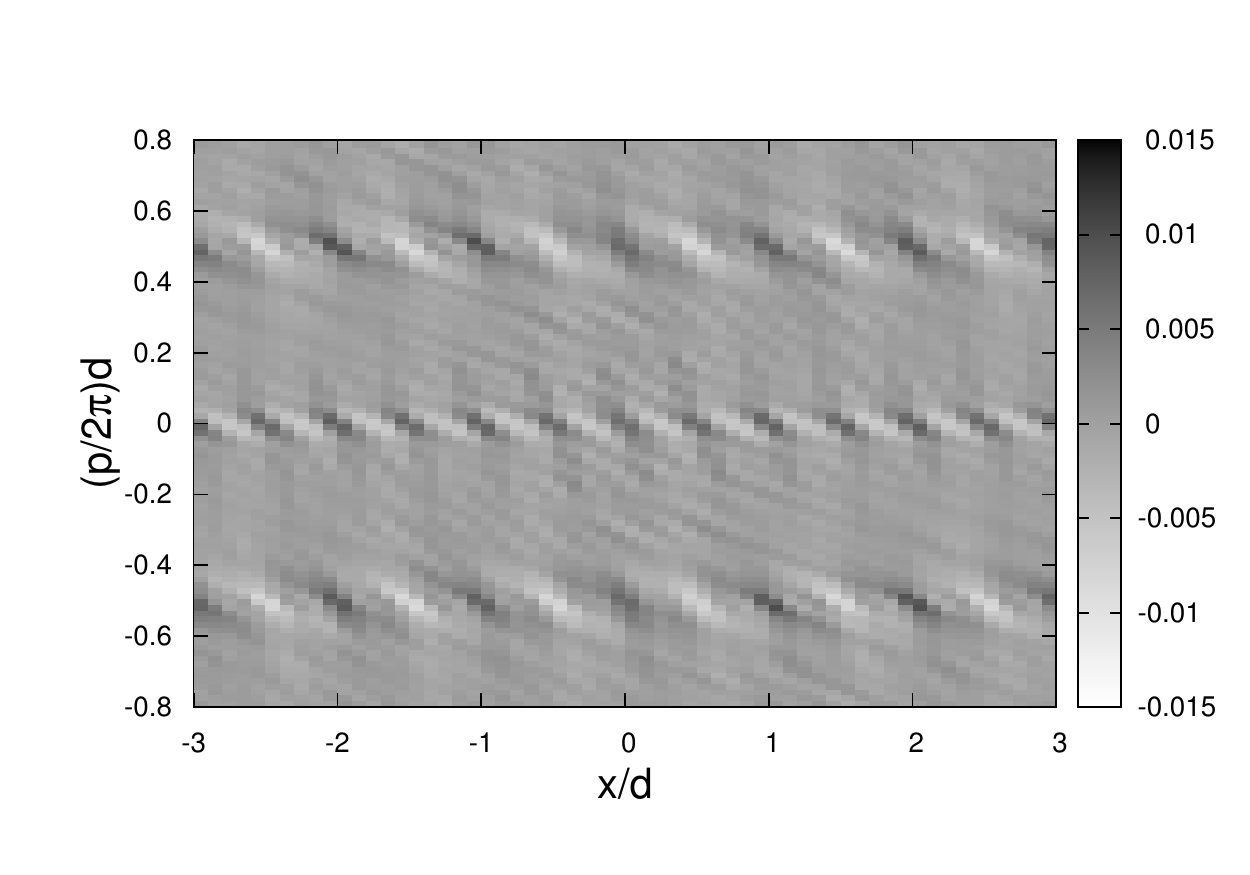}
\includegraphics[scale=0.65]{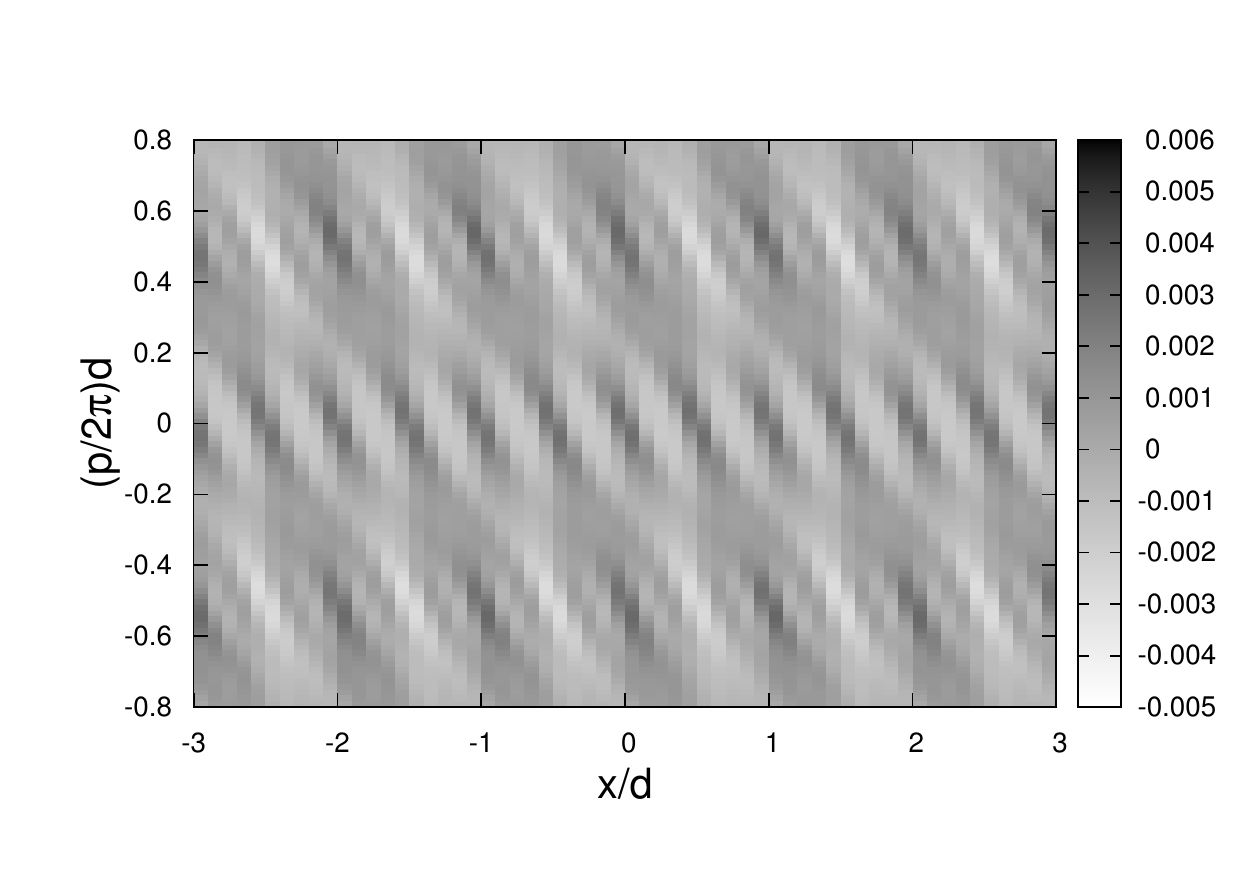}
\caption{(a) Propagation distance $z$ as a function of the rotation angle $\theta$, with distance $z$ scaled by Talbot distance $z_T$. (b) Full reconstruction of WDF, and partial reconstruction of the WDF with rotation angle between [0,$4z_T$] (c) and [0, $z_T$] (d).}
\label{fig:ztheta}
\end{figure}

The WDF is rotated by free space propagation. In the experiment the rotation angle is described by the distance $z$ after the grating:
\begin{equation}
z = \frac{2\pi}{\lambda}\tan\theta.
\label{eq:z_theta}
\end{equation}
Since we use $z$ and $\lambda$ in units of $d$, Eq.~(\ref{eq:z_theta}) becomes $z/z_T=\pi \tan\theta$ which is illustrated in Fig.\ref{fig:ztheta}(a). As a result a rotation of close to $\pi$/2 can be achieved with several Talbot distances $z_T$, although a rotation of exactly $\pi$/2 corresponds to infinite distance. A full reconstruction of the WDF is therefore not feasible, and we now investigate the \emph{partial} reconstruction for limited rotation angles to investigate whether this still leads to negativity of WDF. In Fig.\ref{fig:ztheta}, we compare full reconstruction (b), partial reconstruction of WDF with rotation angle [0,4$z_T$] (c), and for [0,$z_T$] (d). We observe tilting of WDF for partial reconstruction, but the position of the peaks and the negativity still remain. When partially reconstructing the WDF, negative peaks appear when including the displaced self-images, which are truly due to interference. We conclude therefore, these negative peaks in the partially reconstructed WDF show non-classical behavior. In the following, we chose a rotation angles between 0 and $4z_T$ which corresponds to rotation angles between 0 to $\arctan [4/\pi]$ according to Eq.~(\ref{eq:z_theta}). We note the rotation angle depends on the unit length and the corresponding WDF is also expressed as a function of the unit length~\cite{pfau1997partial}.

\subsection{Resolution in x and z}
\begin{figure}
(a)\hspace{8cm} (b)\\
\includegraphics[scale=0.65]{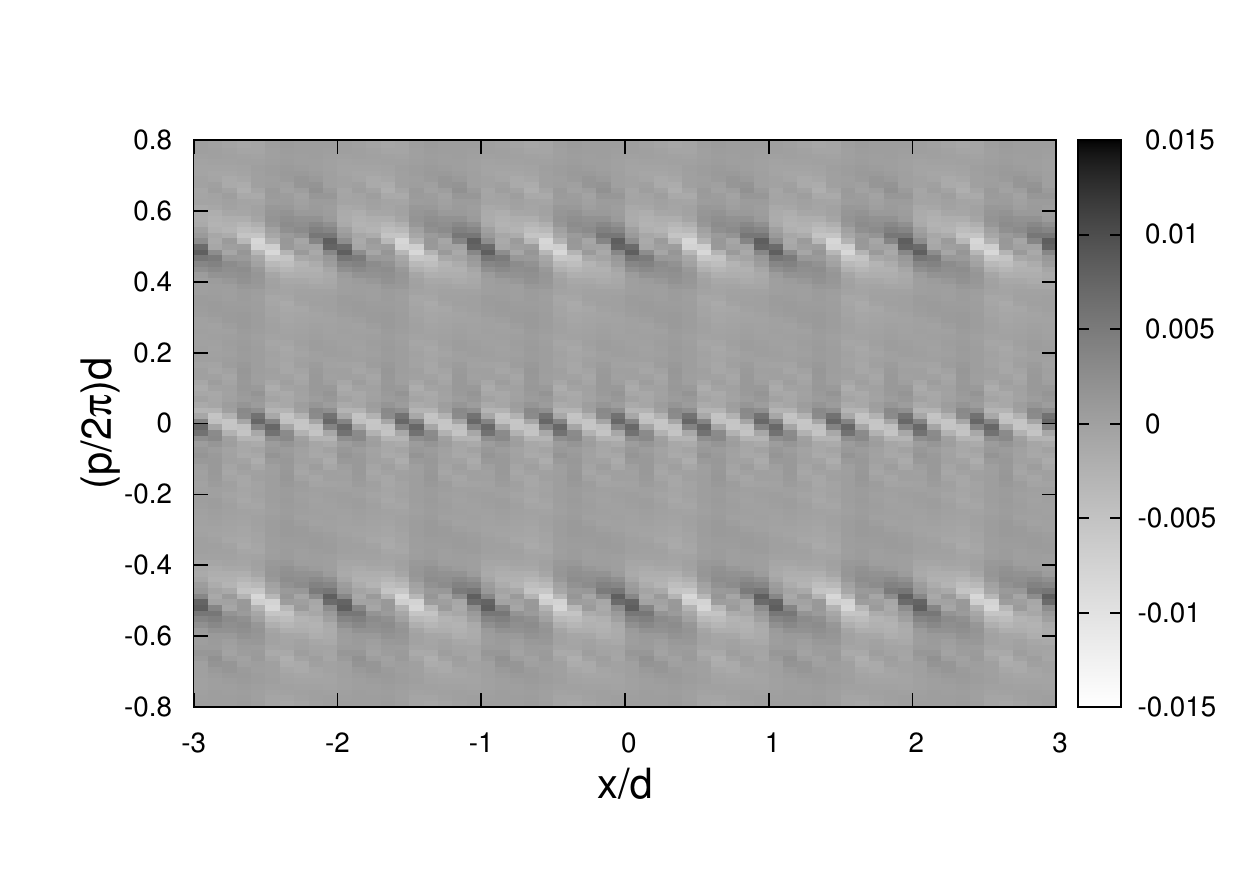}
\includegraphics[scale=0.65]{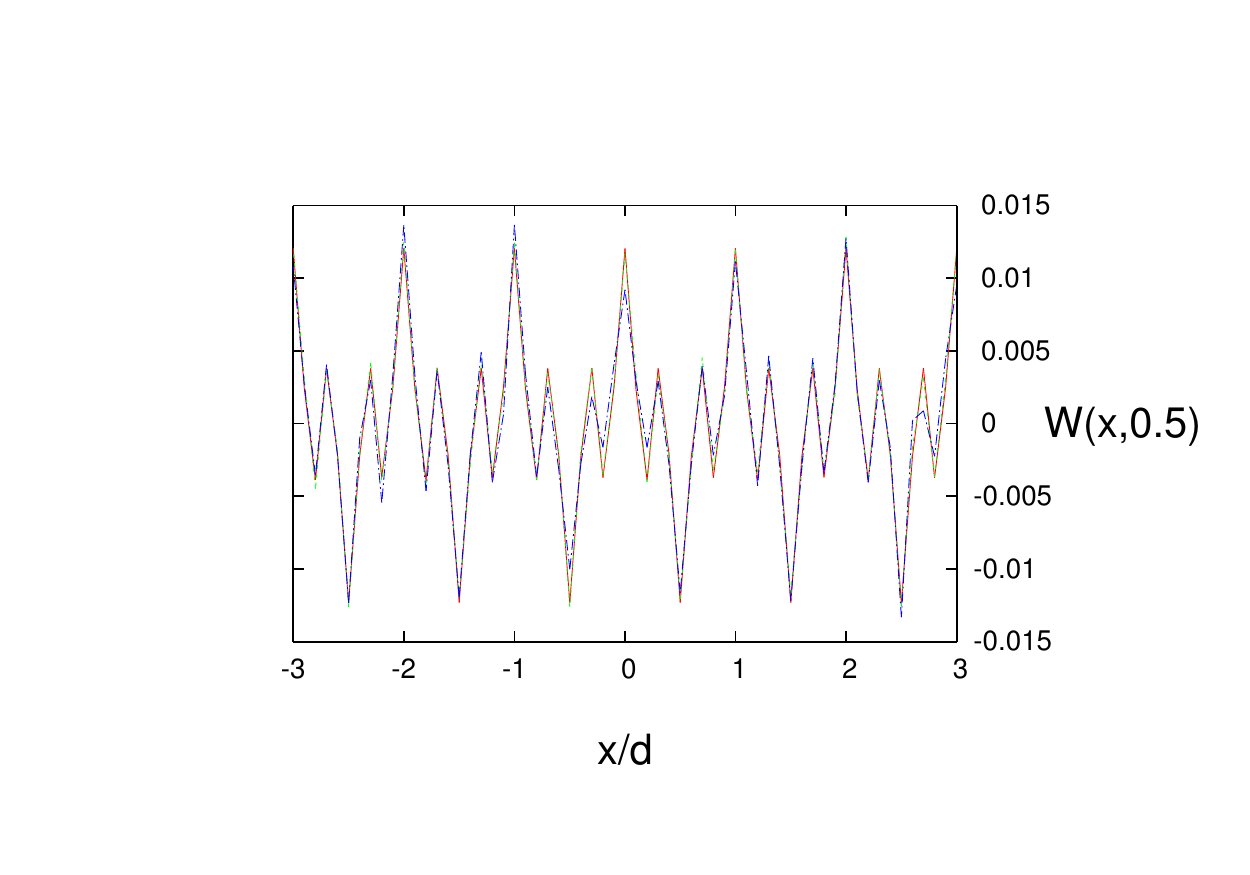}
\caption{Partial reconstruction of WDF between 0 to 4$z_T$ varying resolution of $x$-axis, which corresponds to the spatial resolution of the detector in the experiment. (a) Contour plot for $dx=0.01d$, (b) Cross section plot for WDF at $(p/2\pi)d=0.5d$ for $dx=0.01d$ (red solid line), $dx=0.05d$ (green dotted line), and $dx=0.1d$ (blue dash-dotted line) where $N_\theta=100$ and $f_0=0.3$.}
\label{fig:prwdf_dx}
\end{figure}

In the last section we investigated the range of rotation angles necessary for observing negativity in the WDF. Now, our interest is the dependency on the resolution in $x$- as well as in $z$-direction. The resolution in x-direction corresponds to the spatial resolution of the detector. Simulations are shown in Fig.~\ref{fig:prwdf_dx}, where we compare $dx=0.01d$, $dx=0.05d$, and $dx=0.1d$. We conclude that a resolution of ten measurement points per grating period $d$ (and possibly even less) are sufficient to reconstruct negativity of WDF. This resolution is typically achieved in state-of-the-art molecule interference experiments. However structure of the reconstructed WDF becomes a more pronounced for higher $x$-resolution. We chose $dx=0.1d$ for the following simulations.

\begin{figure}
(a)\hspace{8cm} (b)\\
\includegraphics[scale=0.65]{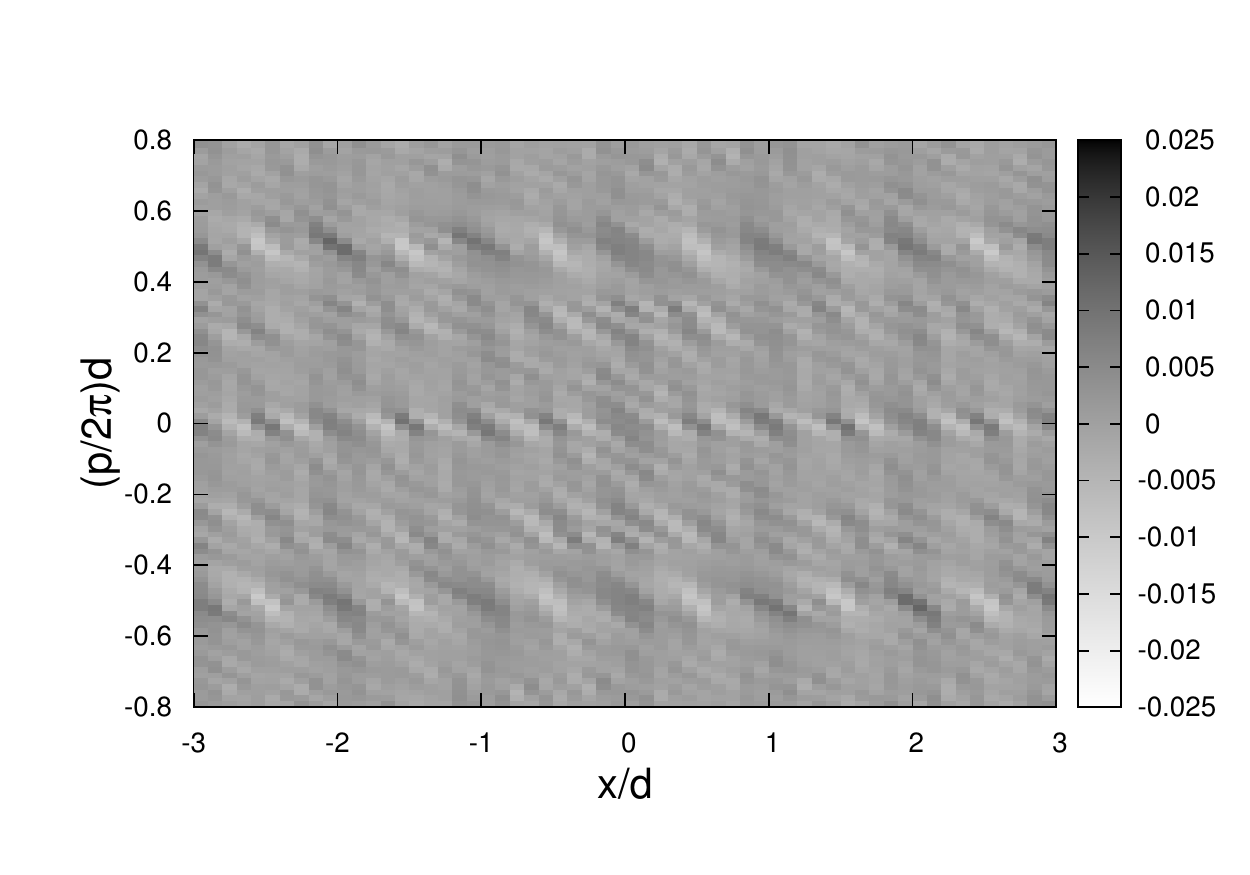}
\includegraphics[scale=0.65]{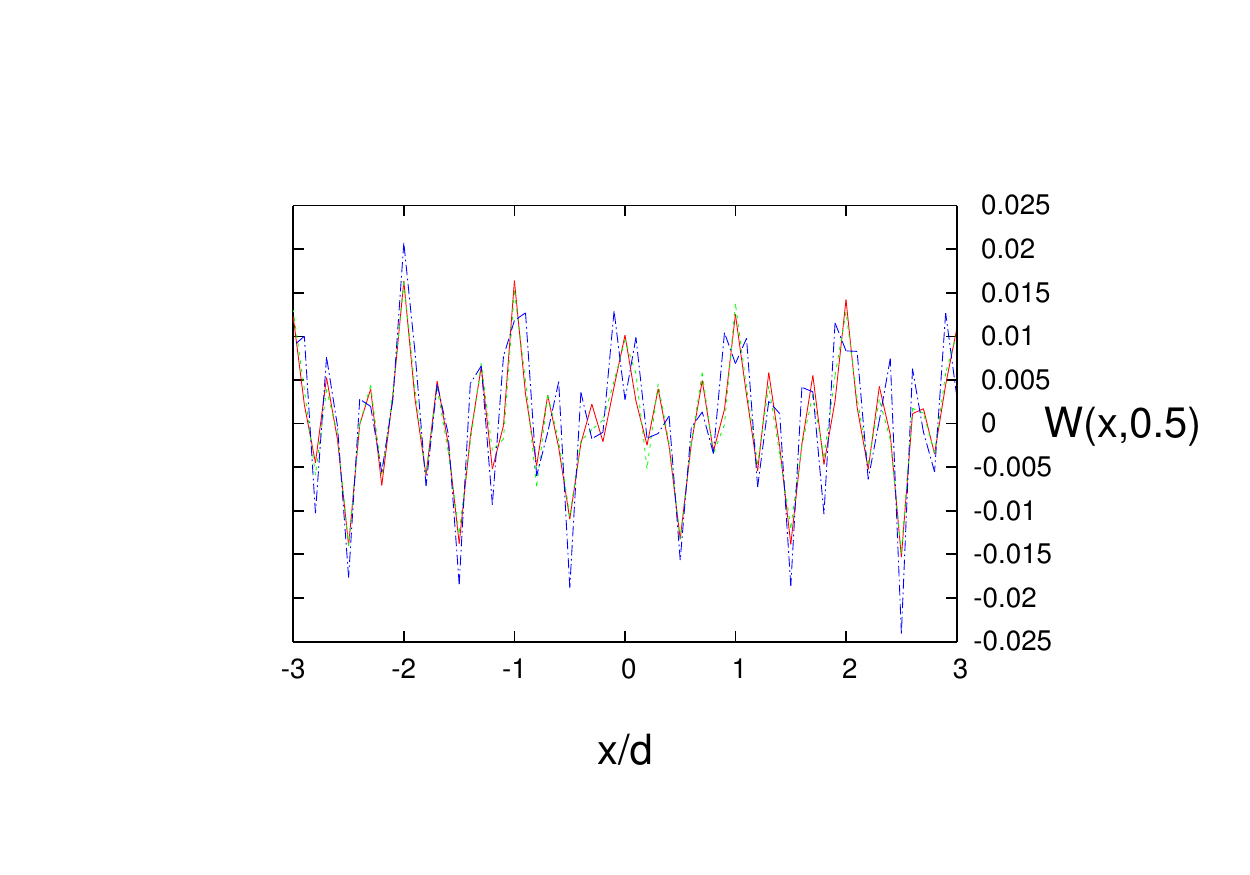}
\caption{Partial reconstruction of WDF from 0 to 4$z_T$ varying the number of rotation angles (a) Contour plot for $N_{\theta}=20$, (b) Cross section plot for WDF at $(p/2\pi)d=0.5$ for $N_\theta=100$(red solid line), $N_\theta=50$ (green dotted line), and $N_\theta=20$ (blue dash-dotted line) where $f_0=0.3$ and $dx=0.1$.}
\label{fig:rwdf_dn}
\end{figure}

We find that an important parameter is the number of rotation angles $N_\theta$ which corresponds to the propagation distance $z$, as discussed in Sec.~\ref{rotationangle}. On the other hand the resolution of the $z$-direction itself is not critical and we show simulations in Fig.~\ref{fig:rwdf_dn} to prove this statement. We keep the the total distance $z$ constant, but change the resolution which is the number of scans within $z$. In Fig.~\ref{fig:rwdf_dn}(a) the reconstructed WDF for $N_{\theta}$=20 within 4$z_T$ is shown. The structure is very similar compared to Fig.~\ref{fig:ztheta}(c), which is over the same $z$ distance. We then vary the resolution further and show cross sections of the contour plot of the WDF at $(p/2\pi)d=0.5$ for $N_{\theta}=20, 50,100$ in Fig.~\ref{fig:rwdf_dn}(b). The periodic negative and positive peaks are almost independent from $N_{\theta}$. Therefore, data taken at the self-image planes are sufficient to reconstruct WDF. That means, when measuring only the self-image planes we need to measure until 4$z_T$ to find negativity in the reconstructed WDF. We note, displaced self images taken at integer multiple of $z_{T}/2$ are crucial to generate negative peaks.

\subsection{Incoherent Source}\label{sec:incoherentsource}
\begin{figure}
(a)\hspace{9cm} (b)\\
\includegraphics[scale=0.65]{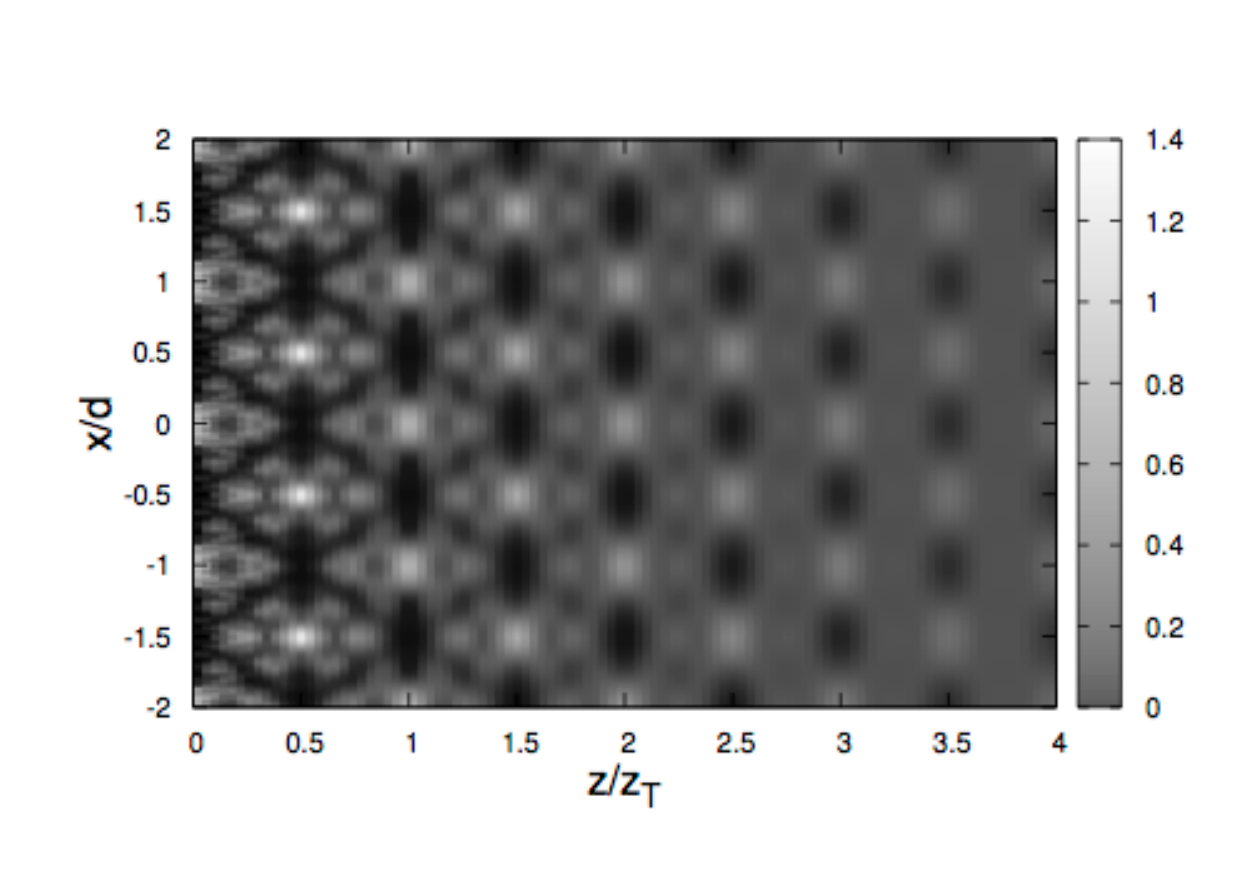}
\includegraphics[scale=0.65]{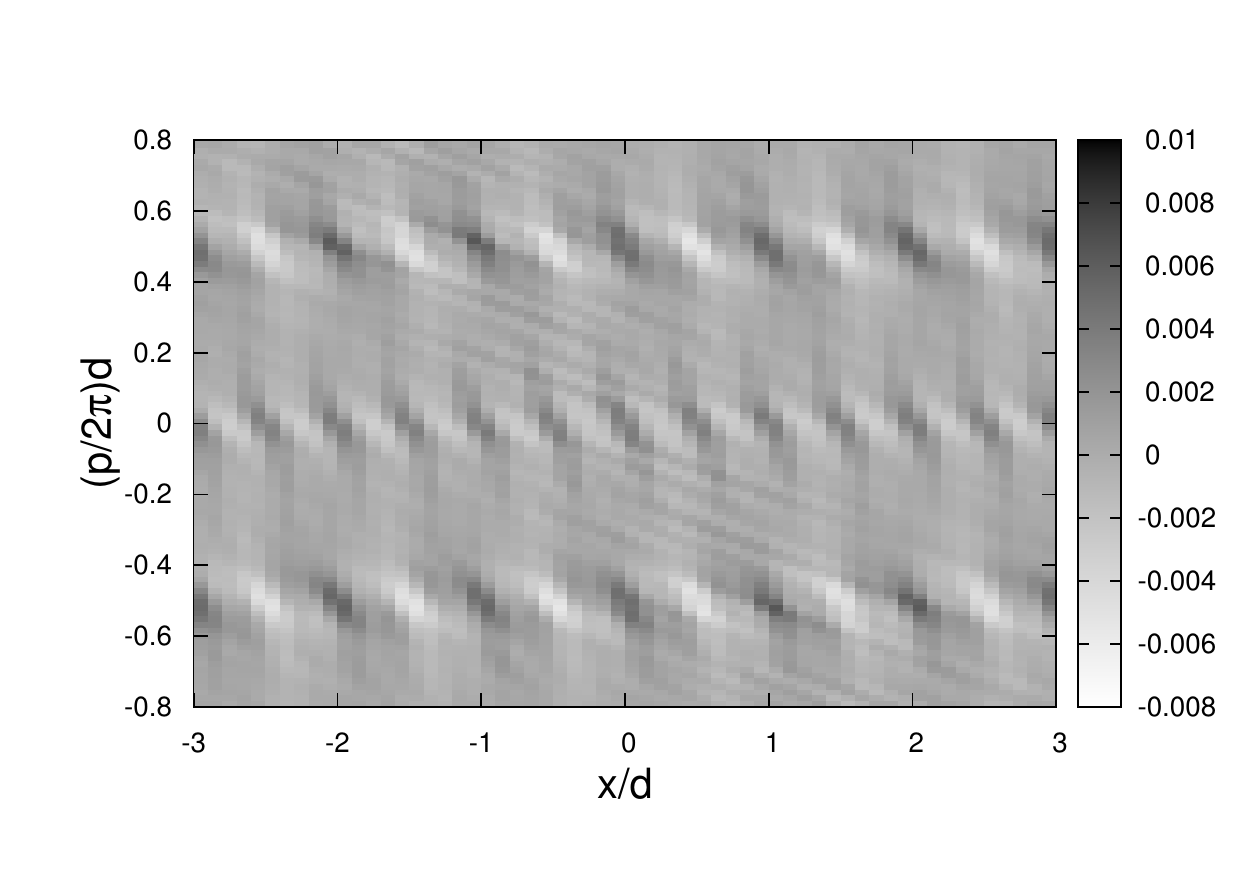} \\
(c)\hspace{9cm} (d)\\
\includegraphics[scale=0.65]{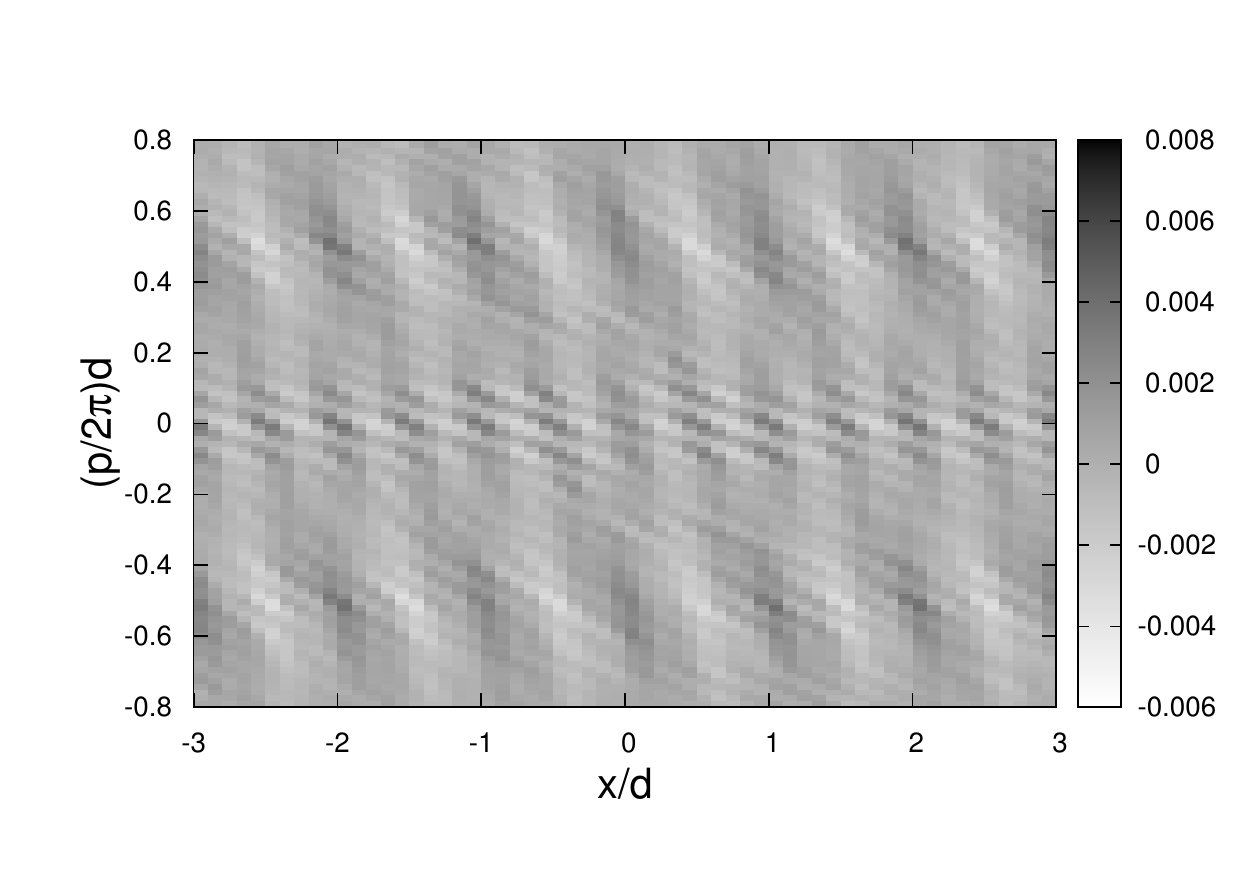}
\includegraphics[scale=0.65]{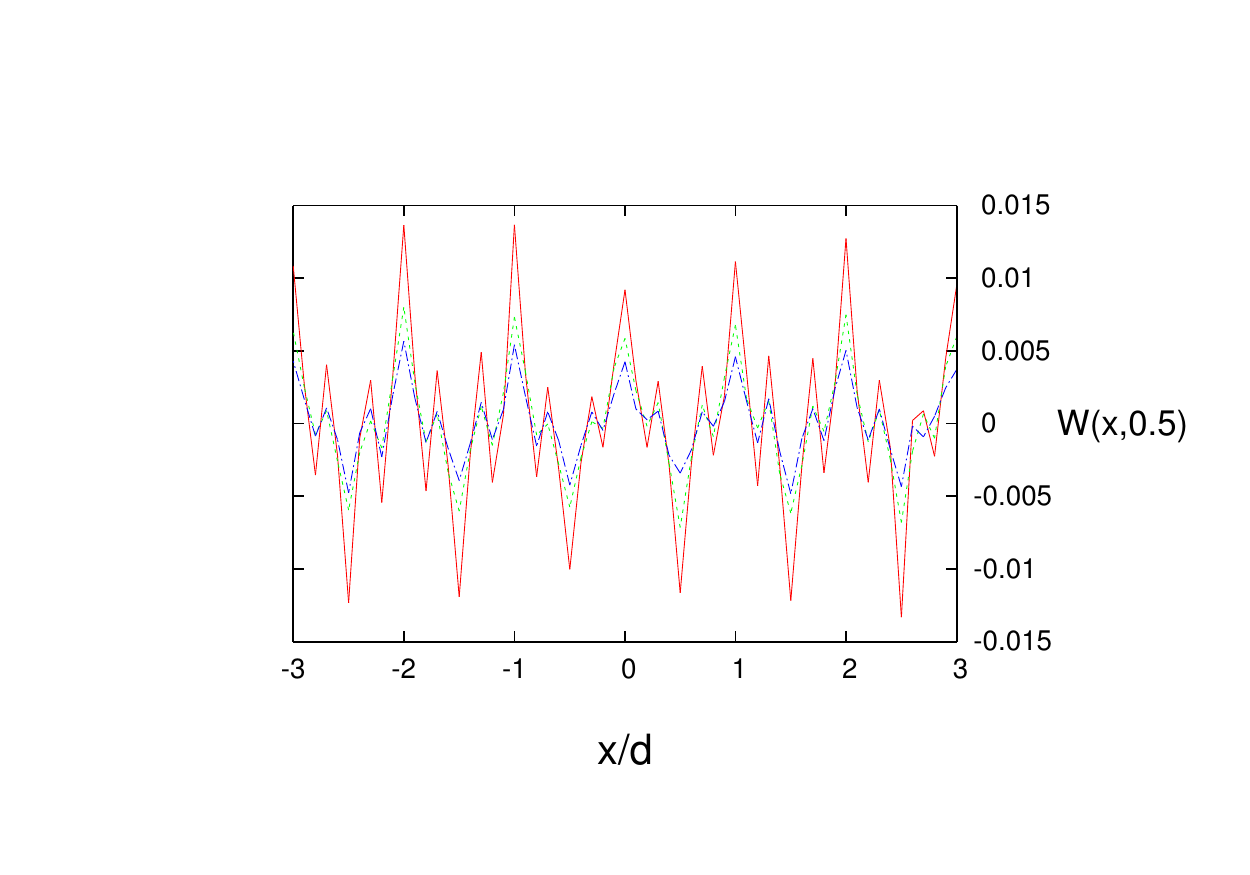}
\caption{Probability distribution (a) and partial reconstruction of the WDF with an incoherent source based on data of $[0, 4z_T]$ with an incident angle of $[\alpha]_{max}=\pi\times10^{-6}$ (b) and $[\alpha]_{max}=2.5\pi\times10^{-6}$(c). (d) The WDF at $(p/2\pi)d=0.5$ for $\alpha_{max}=0$ (red solid line), $\alpha_{max}=\pi\times10^{-6}$ (green dotted line), and $5\pi\times10^{-6}$ (blue dash-dotted line)  where $f_0=0.3, N_\theta=100, dx=0.1d$.}
\label{fig:prwdf_dk}
\end{figure}
Effects of temporal coherence, which correspond to the longitudinal $z$-velocity selection of the molecular beam is taken into account by a reduction of the visibility for a given distance $z$ in Sec.~(\ref{sec:visibility}). Here, we discuss the effect of an incoherent source with respect to spatial coherence of the molecular matter waves. This corresponds to the transverse $x$-velocity selection by collimation of the molecular beam. Assuming an incident collimation of angle $\alpha$ (see Fig.~(\ref{fig:talbot})), the wave function is found by averaging over all incident angles $\alpha$:
\begin{equation}
\psi(x,z) = \left<\sum_{n=-\infty}^{\infty} A_n \exp[{i(k_{\alpha}+n\frac{2\pi}{d}x)+i(k-\frac{{(k_{\alpha}+n\frac{2\pi}{d}x)}^2}{2k})z}]\right>_{\alpha} ,
\end{equation}
where $k=({k_z}^2+{(k_{\alpha}+n\frac{2\pi}{d}x)}^2)^{1/2}$ and $k_{\alpha} = k \sin(\alpha)$. The variation of the beam collimation is implemented mathematically by the standard deviation of a Gaussian distribution of $\sigma=0.1\alpha_{max}$, with $\alpha_{max}$ is the total range of incident collimation angles. The wave function is then averaged between
[$-\alpha_{max}/2$ , $\alpha_{max}/2$] with Gaussian weight. The Talbot effect can be observed in the limit $\alpha <d/z_T$, which means that the diffraction is dominating collimation. The probability distribution is plotted in Fig.~\ref{fig:prwdf_dk}(a) for incident angle $\alpha_{max}=\pi\times10^{-6}$, which is easy to achieve in the experiment by collimation at the fist grating in a Talbot and Talbot-Lau interferometer. We observe that interference contrast is washed out rapidly for increasing $z$. We reconstruct WDF as shown in Fig.~\ref{fig:prwdf_dk}(b) and it still shows negative peaks between positive peaks. However, we observe a significant lost of contrast of WDF for increased collimation angle ${\alpha}_{max}$, as shown in Fig.~\ref{fig:prwdf_dk}(c) and (d). Since the interference pattern are very sensitive to the spatial coherence, the Talbot-Lau interferometer configuration is used in experiments. Here, an additional grating placed in front of the diffraction grating increases the spatial coherence of the matter waves and Talbot images can be well revived even with a molecular beam source of relatively poor spatial coherence~\cite{clauser1992}.

\subsection{Finite Gratings}
\begin{figure}
(a)\hspace{9cm}(b) \\
\includegraphics[scale=0.65]{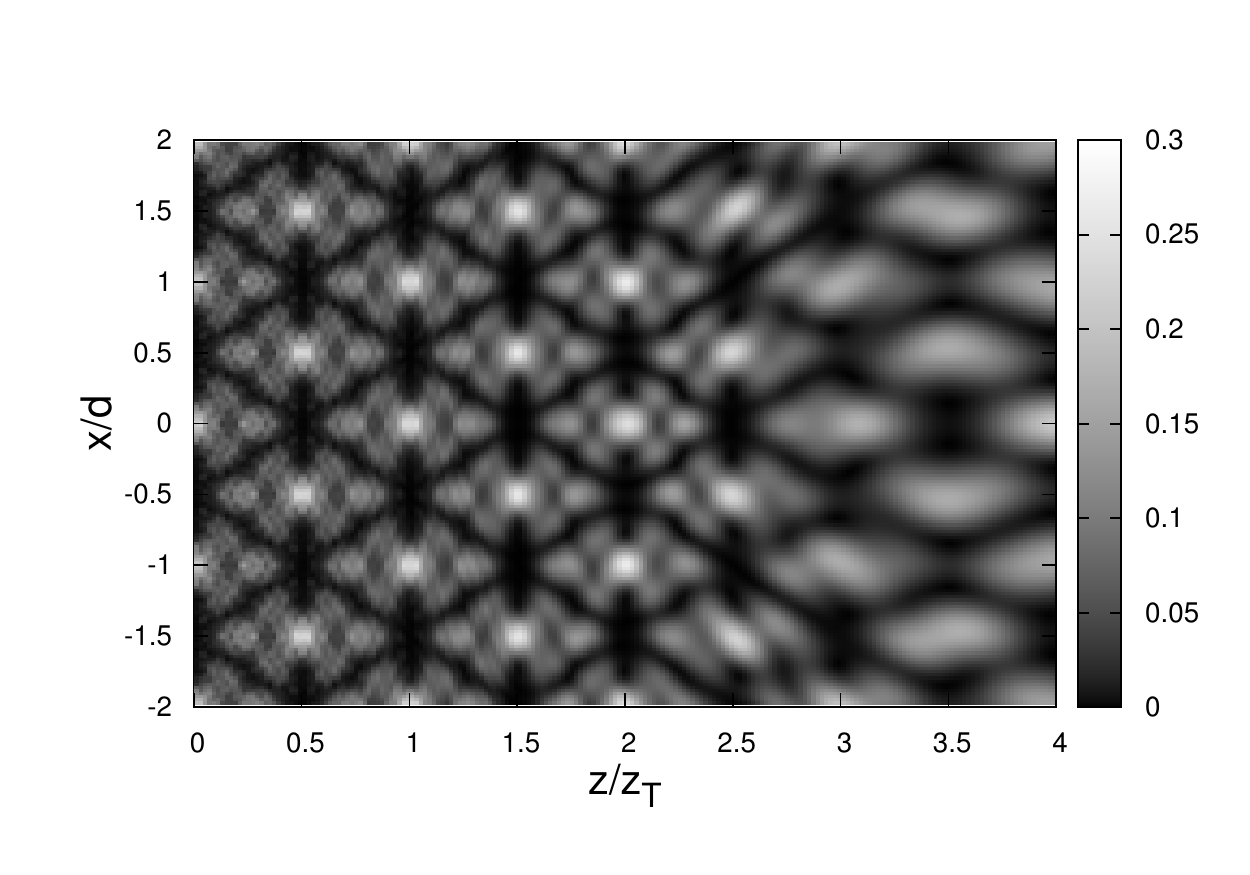}
\includegraphics[scale=0.65]{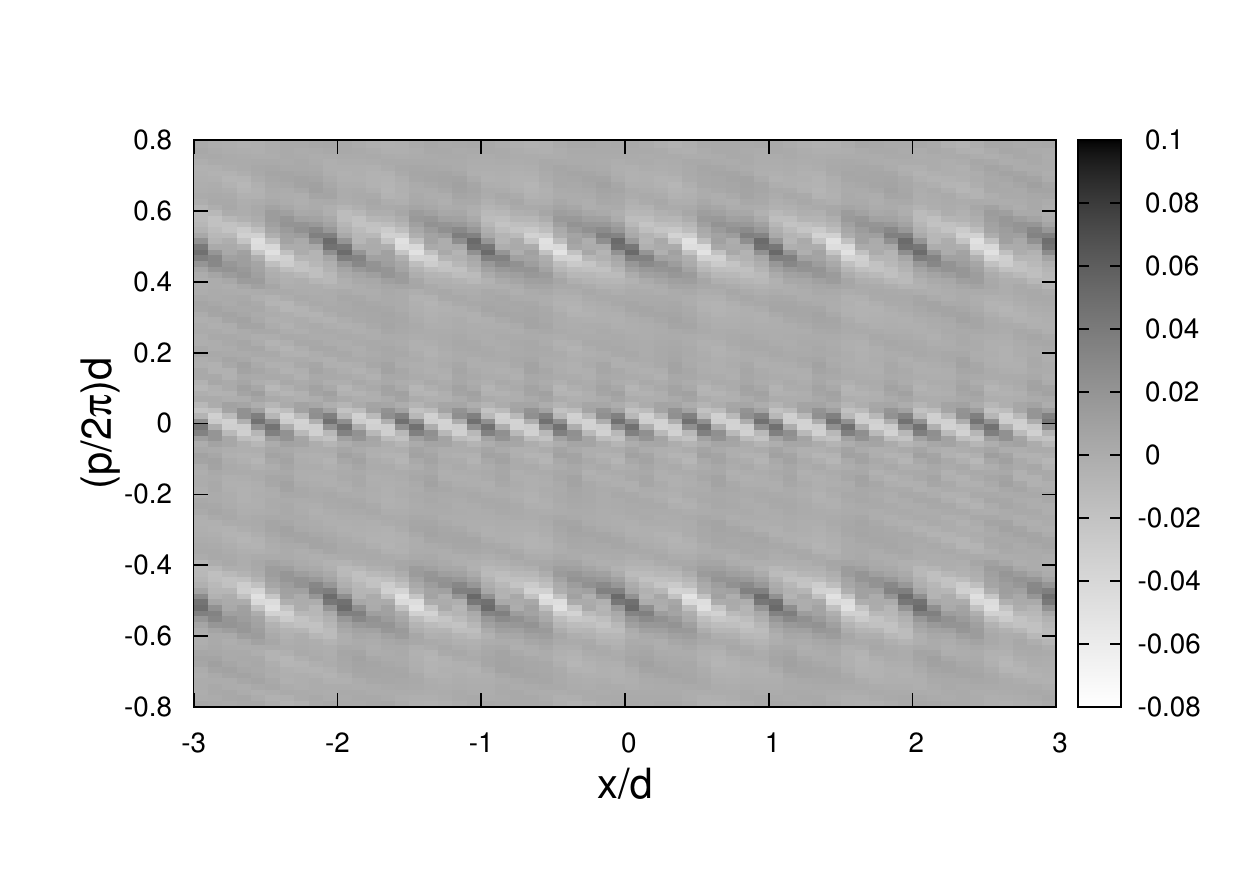}
\caption{Probability distribution (a) and partial reconstruction of the WDF (b) for finite number of slits, $N_s$=10 for $[0,4 z_T]$ where $N_\theta=100, f_0=0.3, dx=0.1$.}
\label{fig:finitepd}
\end{figure}

The grating was assumed to have an infinite number of slits until now. Experimentally, the effective number of involved slits is finite, while sufficiently large to observe the Talbot effect. Talbot self-images, however, are washed out after some multiple of Talbot distance $z_T$ as an effect of the finite number of slits~\cite{Hornberger2012}. To study this effect quantitatively, we define the wave function for a finite number of slits as:
\begin{equation}
\psi(x)=t_c(x)\otimes \sum_{n=-N_s}^{n=N_s}\delta(x-nd),
\end{equation}
with 2$N_s$ slits. The wave function after the grating in free space propagation is then given by:
\begin{equation}
\psi(x,z)=t_c(x)\otimes \sum_{n=-N_s}^{n=N_s} e^{i\frac{\pi}{\lambda z}(x-nd)^2}.
\end{equation}
We simulate density distribution and WDF for various number of slits. Contrast in the density distribution is still visible in Fig.~\ref{fig:finitepd}(a) for as few as ten grating slits $N_s$ contributing to diffraction for up to three Talbot distances and we can still reconstruct WDF with negativity in Fig.~\ref{fig:finitepd}(b). Both, the visibility of positive and negative peaks of WDF as well as the distance (number of $z_T$), where near-field pattern can be seen in the density distribution, increase as the number of slits increases.

\subsection{Van der Waals Interaction}
\begin{figure}
(a)\hspace{9cm}(b) \\
\includegraphics[scale=0.65]{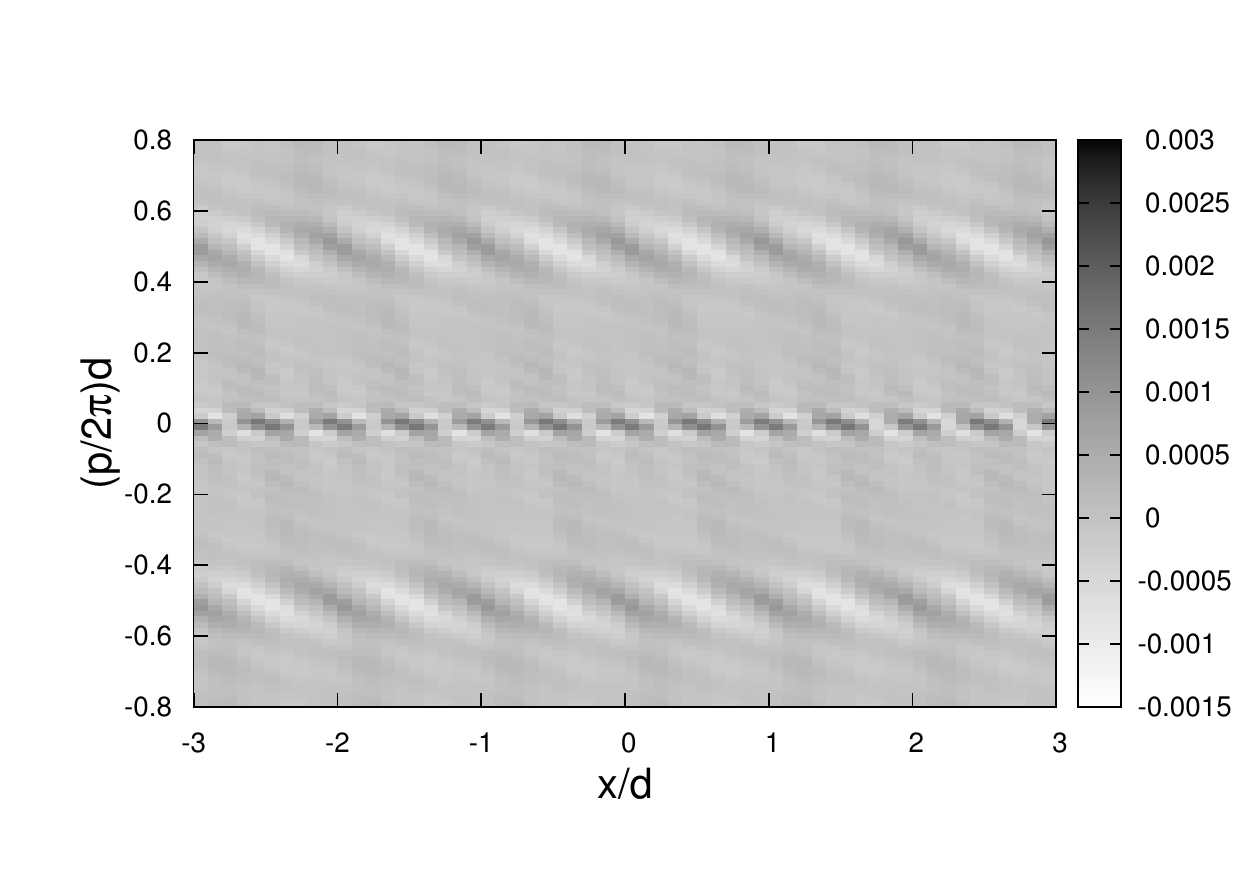}
\includegraphics[scale=0.65]{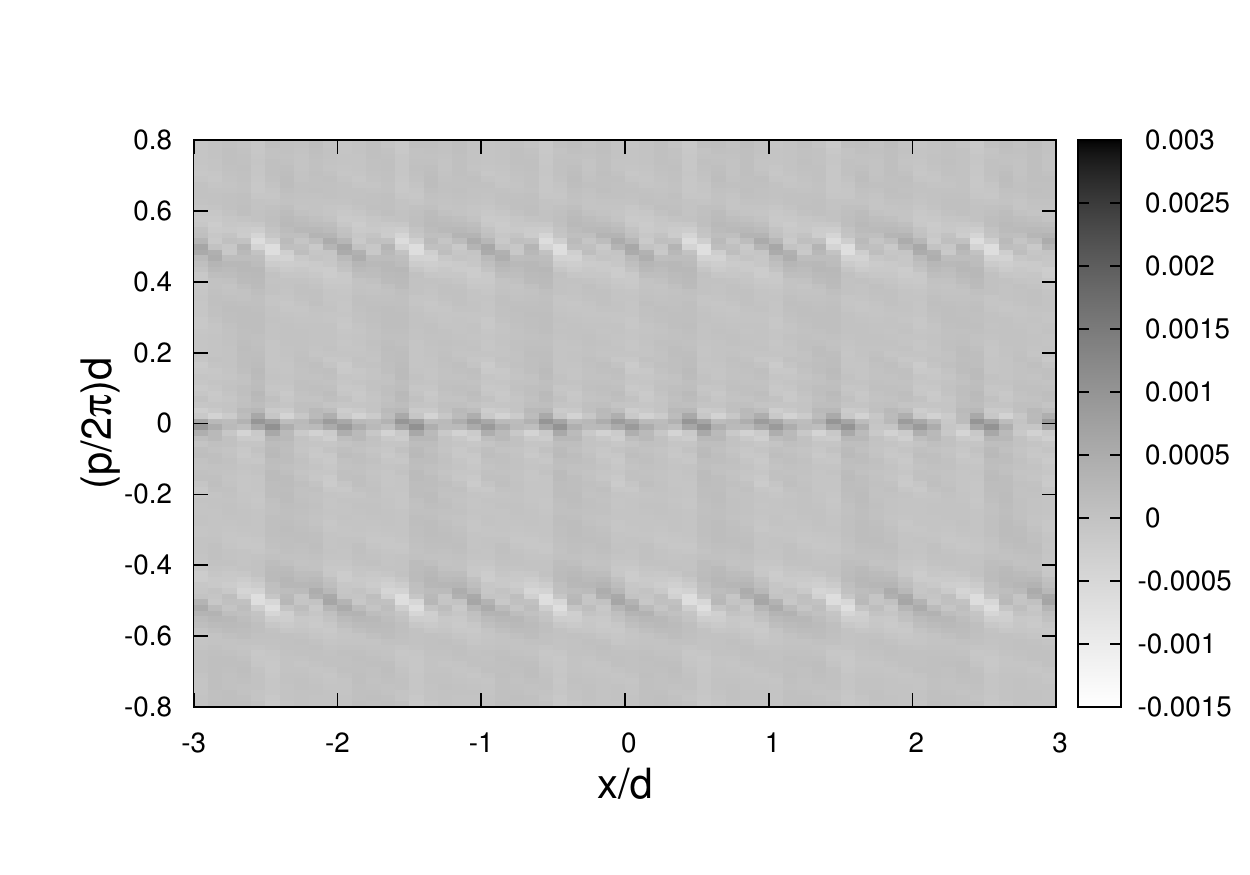} \\
\caption{The partial reconstructed WDF for $[0,4z_T]$ without van der Waals (vdW) interaction (a) and with vdW interaction (b) where open fraction $f_0=0.44$, $dx=0.1d, N_\theta=80$.}
\label{fig:rwdfvdw}
\end{figure}

Here we investigate the effect of the interaction between the grating and molecules on the WDF reconstruction. Dispersion forces, such as van der Waals (vdW) forces in the short range limit (on the order of 100~nm between molecule and grating wall) and Casimir-Polder (CP) forces in the long range limit (for larger distances than some 100~nm), are known to affect the molecule interference pattern~\cite{nimmrichter2008theory}, if the gratings are realized from material structures made of metals (gold, Au) or semiconductors (silicon nitride, SiN$_x$). Dispersion forces will therefore have a large effect on the quantum carpet structure and we have to test the respective limits of WDF reconstruction. We implement the grating wall molecule interaction as a phase term $\phi(x)$ to transmission function $t'_c(x)=t_c(x)\phi(x)$ in Eq.~(\ref{eq:combf}):
\begin{equation}
\phi(x)=e^{\frac{im}{\hbar p}\int^{\infty}_{-\infty}dz\,V(x,z)},
\end{equation}
with the mass $m$ of the molecule and the VdW potential $V=-C_3/x^3$, which we exclusively discuss here for simplicity. The $1/x^3$ scaling of the potential with distance $x$ is typical for a pointlike particle in front of a surface. The vdW interaction constant $C_3$ is depending on the dielectric properties of the molecule and the grating over the full electromagnetic spectrum. For our simulations we use $C_3$=10~meV~nm$^3$ for a fullerene C$_{60}$ molecule close to a gold surface in agreement with literature~\cite{nimmrichter2008theory}. Fig.~\ref{fig:rwdfvdw} shows reconstructed WDF (a) without and (b) with vdW interaction and it shows a small change in visibility and structure. We note, that density distribution (quantum carpet) and WDF with vdW are very similar to simulations with smaller open fraction $f_0=0.1$ without vdW. This is in agreement with the known effect that the always attractive vdW interaction effectively reduces the slit width of the diffraction grating~\cite{brezger2002}. The reduced visibility in the density distribution and the reduced negativity of WDF in Fig.~\ref{fig:rwdfvdw} (b) are explained by dephasing.

\subsection{Visibility}\label{sec:visibility}
\begin{figure}
(a)\hspace{9cm}(b) \\
\includegraphics[scale=0.65]{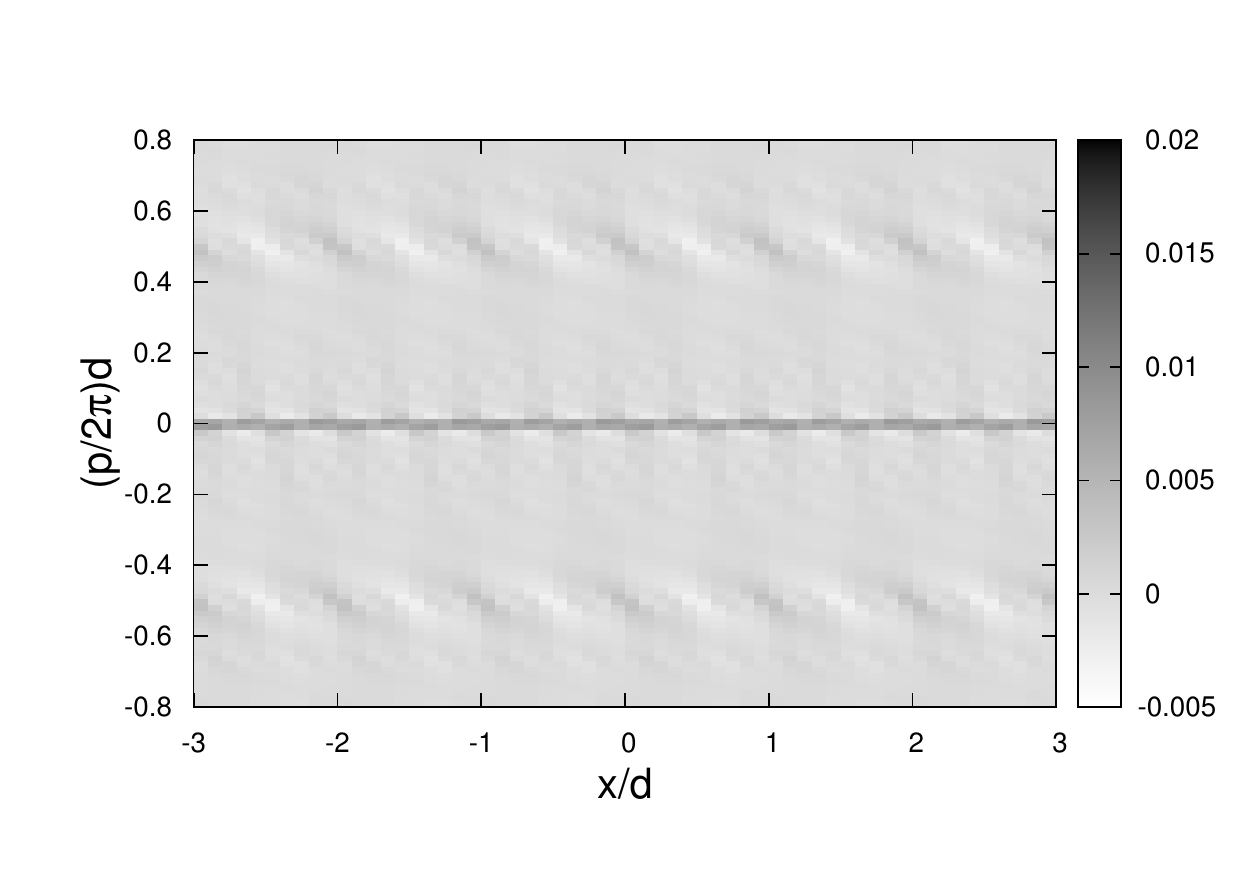}
\includegraphics[scale=0.65]{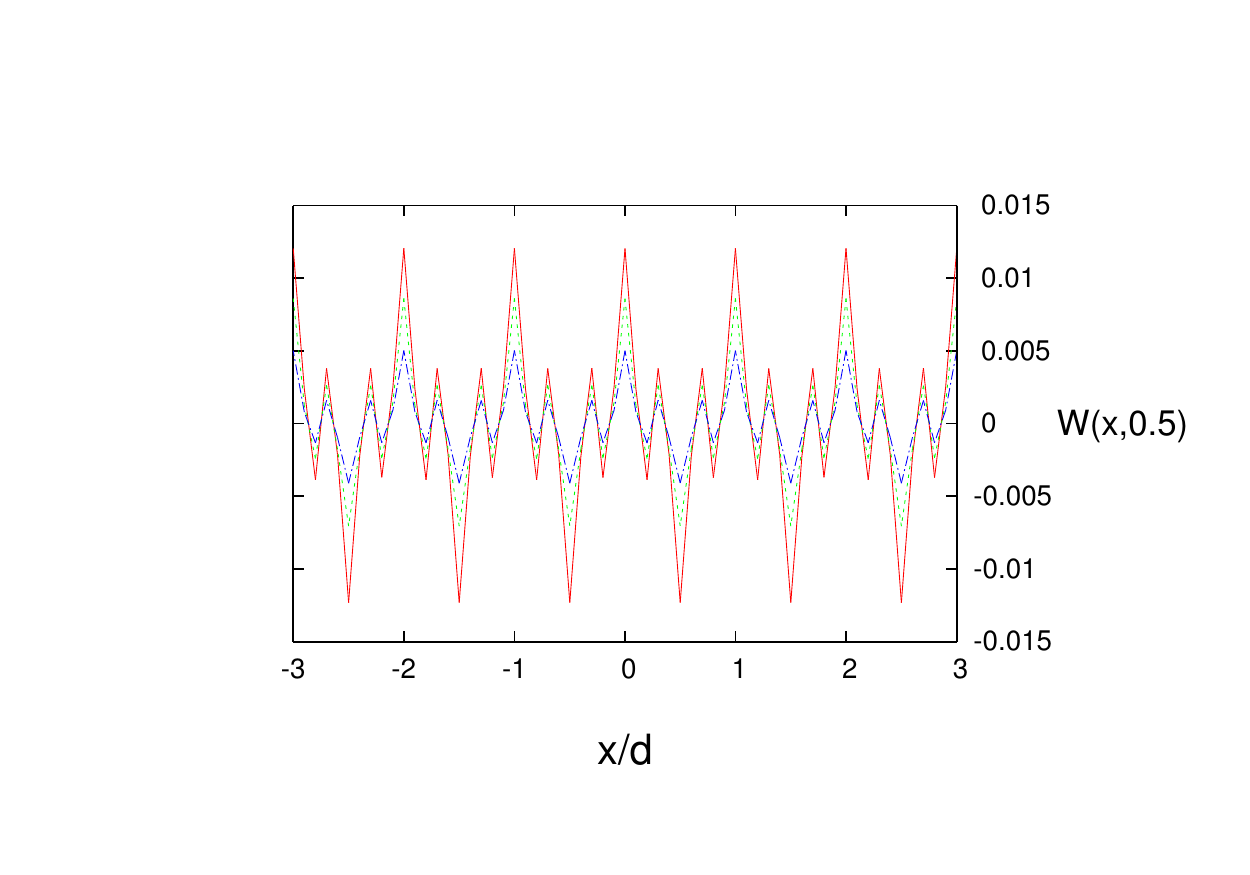} \\
\caption{(a) Contour plot of partially reconstructed WDF for $[0,4z_T]$ with visibility $\mathcal{V}=$0.5, and (b) cross section plot of the WDF at $(p/2\pi)d=0.5$ for $\mathcal{V}=1$ (red solid line), $\mathcal{V}=0.75$ (green dotted line), and $\mathcal{V}=0.5$ (blue dash-dotted line) where $N_\theta=100, f_0=0.3, dx=0.1d$.}
\label{fig:rwdfvis}
\end{figure}

Another parameter which will be investigated here is the visibility $\mathcal{V}=(I_{max}-I_{min})/(I_{max}+I_{min})$, where $I$ is the intensity of the quantum carpet structure illustrated by the gray scale in the plots. The visibility is the most sensitive parameter of the pattern an is easily influenced by different effects. For instance a low velocity selection of the molecular beam resulting in a low longitudinal or temporal coherence of the matter wave reduces the visibility of the carpet. The negativity of WDF is therefore also reduced. Furthermore as already mentioned in Sec.~\ref{sec:incoherentsource}, the Talbot effect is experimentally usually observed in the Talbot-Lau interferometer. The main difference between Talbot and Talbot-Lau from the theoretical point of view is a reduction in interference fringe visibility for the TLI. This means our discussion on WDF reconstruction in the Talbot regime can be also extended to the Talbot-Lau regime, if we investigate the effect of visibility reduction. To include the visibility in the probability distribution (quantum carpet) we modify the density distribution $P(x.z)$ to be: $I_{min}+(1-I_{min})P(x,z)$, with $I_{min}=(1-\mathcal{V})/(1+\mathcal{V})$. We reconstruct WDF and plot in Fig.~\ref{fig:rwdfvis}(a), vary the visibility and find that with a visibility of $\mathcal{V}$=0.5 we can still identify negative parts of the WDF although the contrast is reduced as shown in Fig.~\ref{fig:rwdfvis}(b). The analysis of visibility by WDF reconstruction will become a useful tool to investigate decoherence effects and mechanisms in molecule quantum optics.

\section{Conclusion}
We have numerically analyzed the partial Wigner distribution function (WDF) reconstruction of near-field optical quantum carpets for free space propagation of the wave. In our study, we considered all the major experimental inefficiencies. Most important, we find that negativity can be observed in the reconstructed Wigner distribution function in the Talbot and Talbot-Lau regime under such realistic conditions with today's technologies. All investigated parameter and effects keep promise for the realization of molecular quantum optics tomography. Important to reconstruct WDF is to collect data of quantum carpets at half-integer multiple of Talbot distances. The Talbot regime is important for molecule  interferometry as the conceptually simpler double slit and far-field interference are much harder to perform due to experimental difficulties. The here presented tomography of the motional quantum state will become an important analytic tool in molecule quantum optics as it gives a handle to directly detect the superposition state of centre of mass motion. Furthermore WDF reconstruction may be used to investigate wave function dephasing effects such as van der Waals or Casimir-Polder interactions of the particles as well as to study decoherence effects which reduce the visibility~$\mathcal{V}$. Further work needs to conduct experimental tomography of the discussed quantum state, while the main challenge lasts to generate an intense beam of large particles while keeping sufficient coherence, which requires high phase-space density. The here described tomographic analysis is also applicable to investigate the quantum superposition of the centre of mass motion of even larger nanoparticle and cluster as well as for electromagnetic waves such as X-rays~\cite{Gaffney08062007}.

\section*{Acknowledgements}
This work has been financially supported by the University of Southampton and the Foundational Questions Institute (FQXi).

\section*{References}

\bibliographystyle{unsrt}
\bibliography{references}

\begin{thebibliography}{10}

\bibitem{nimmrichter2011testing}
S.~Nimmrichter, K.~Hornberger, P.~Haslinger, and M.~Arndt.
\newblock Testing spontaneous localization theories with matter-wave
  interferometry.
\newblock {\em Physical Review A}, 83(4):043621, 2011.

\bibitem{romero2011optically}
O.~Romero-Isart, A.C. Pflanzer, M.L. Juan, R.~Quidant, N.~Kiesel,
  M.~Aspelmeyer, and J.I. Cirac.
\newblock {Optically Levitating Dielectrics in the Quantum Regime: Theory and
  Protocols}.
\newblock {\em Physical Review A}, 83(1):013803, 2011.

\bibitem{Gerlich2011quan}
S.~Gerlich, S.~Eibenberger, M.~Tomandl, S.~Nimmrichter, K.~Hornberger, P.~J.
  Fagan, J.~T\"uxen, M.~Mayor, and M.~Arndt.
\newblock {Quantum interference of large organic molecules}.
\newblock {\em Nature Communications}, 2:263, 2011.

\bibitem{wigner1932quantum}
E.~Wigner.
\newblock On the quantum correction for thermodynamic equilibrium.
\newblock {\em Physical Review}, 40(5):749, 1932.

\bibitem{schleich2001}
W.P. Schleich.
\newblock {\em Quantum Optics in Phase Space}.
\newblock Wiley-VCH, Berlin, 2001.

\bibitem{bertrand1987tomographic}
J.~Bertrand and P.~Bertrand.
\newblock A tomographic approach to wigner's function.
\newblock {\em Foundations of Physics}, 17(4):397--405, 1987.

\bibitem{Vogel1989}
K.~Vogel and H.~Risken.
\newblock Determination of quasiprobability distributions in terms of
  probability distributions for the rotated quadrature phase.
\newblock {\em Phys. Rev. A}, 40:2847--2849, 1989.

\bibitem{leonhardt1997measuring}
U.~Leonhardt.
\newblock {\em Measuring the Quantum State of Light}.
\newblock Cambridge University Press, Cambridge, 1997.

\bibitem{lundeen2008tomography}
J.S. Lundeen, A.~Feito, H.~Coldenstrodt-Ronge, KL~Pregnell, C.~Silberhorn, T.C.
  Ralph, J.~Eisert, M.B. Plenio, and I.A. Walmsley.
\newblock Tomography of quantum detectors.
\newblock {\em Nature Physics}, 5(1):27--30, 2008.

\bibitem{schmied2011tomographic}
R.~Schmied and P.~Treutlein.
\newblock Tomographic reconstruction of the wigner function on the bloch
  sphere.
\newblock {\em New Journal of Physics}, 13:065019, 2011.

\bibitem{leibfried1996experimental}
D.~Leibfried, D.M. Meekhof, B.E. King, C.~Monroe, W.M. Itano, and D.J.
  Wineland.
\newblock Experimental determination of the motional quantum state of a trapped
  atom.
\newblock {\em Physical review letters}, 77(21):4281--4285, 1996.

\bibitem{kurtsiefer1997measurement}
C.~Kurtsiefer, T.~Pfau, and J.~Mlynek.
\newblock Measurement of the wigner function of an ensemble of helium atoms.
\newblock {\em Nature}, 386(6621):150--153, 1997.

\bibitem{janicke1995tomography}
U.~Janicke and M.~Wilkens.
\newblock Tomography of atom beams.
\newblock {\em Journal of Modern Optics}, 42(11):2183--2199, 1995.

\bibitem{romero2011}
O.~Romero-Isart, A.C. Pflanzer, M.L. Juan, R.~Quidant, N.~Kiesel,
  M.~Aspelmeyer, and J.I. Cirac.
\newblock Optically levitating dielectrics in the quantum regime: Theory and
  protocols.
\newblock {\em Physical Review A}, 83(1):013803, 2011.

\bibitem{Vanner2011}
M.R. Vanner, I.~Pikovski, G.D. Cole, M.S. Kim, C.~Brukner, K.~Hammerer, G.J.
  Milburn, and M.~Aspelmeyer.
\newblock Pulsed quantum optomechanics.
\newblock {\em PNAS}, 108:16182, 2011.

\bibitem{steffen2006measurement}
M.~Steffen, M.~Ansmann, R.C. Bialczak, N.~Katz, E.~Lucero, R.~McDermott,
  M.~Neeley, E.M. Weig, A.N. Cleland, and J.M. Martinis.
\newblock Measurement of the entanglement of two superconducting qubits via
  state tomography.
\newblock {\em Science}, 313(5792):1423, 2006.

\bibitem{nimmrichter2008theory}
S.~Nimmrichter and K.~Hornberger.
\newblock {Theory of Near-Field Matter-Wave Interference beyond the Eikonal
  Approximation}.
\newblock {\em Physical Review A}, 78(2):023612, 2008.

\bibitem{brezger2002}
B.~Brezger, L.~Hackerm\"uller, S.~Uttenthaler, J.~Petschinka, M.~Arndt, and
  A.~Zeilinger.
\newblock {Matter-wave interferometer for large molecules}.
\newblock {\em Physical review letters}, 88(10):100404, 2002.

\bibitem{clauser1992}
J.F. Clauser and M.W. Reinsch.
\newblock {New theoretical and experimental results in Fresnel optics with
  applications to matter-wave and X-ray interferometry}.
\newblock {\em Applied Physics B: Lasers and Optics}, 54(5):380--395, 1992.

\bibitem{Hornberger2012}
K.~Hornberger, S.~Gerlich, P.~Haslinger, S.~Nimmrichter, and M.~Arndt.
\newblock \textit{Colloquium}: Quantum interference of clusters and molecules.
\newblock {\em Rev. Mod. Phys.}, 84:157--173, 2012.

\bibitem{friesch2000}
O.M. Friesch, I.~Marzoli, and W.P. Schleich.
\newblock Quantum carpets woven by wigner functions.
\newblock {\em New Journal of Physics}, 2:4, 2000.

\bibitem{berry2001}
M.V. Berry, I.~Marzoli, and W.~Schleich.
\newblock Quantum carpets, carpets of light.
\newblock {\em Physics World}, 14(6):39--44, 2001.

\bibitem{case2009realization}
W.B. Case, M.~Tomandl, S.~Deachapunya, and M.~Arndt.
\newblock Realization of optical carpets in the talbot and talbot-lau
  configurations.
\newblock {\em Optics Express}, 17(23):20966--20974, 2009.

\bibitem{clauser2008}
J.F. Clauser and J.P. Dowling.
\newblock Factoring integers with young's n-slit interferometer.
\newblock {\em Physical Review A}, 53(6):4587--4590, 2008.

\bibitem{schleich2008}
W.~Schleich.
\newblock Factorisation of numbers, schr{\"o}dinger cats and the riemann
  hypothesis.
\newblock In {\em Frontiers in Optics}. Optical Society of America, 2008.

\bibitem{gilowski2008gauss}
M.~Gilowski, T.~Wendrich, T.~M{\"u}ller, C.~Jentsch, W.~Ertmer, EM~Rasel, and
  WP~Schleich.
\newblock Gauss sum factorization with cold atoms.
\newblock {\em Physical Review Letters}, 100(3):30201, 2008.

\bibitem{carnal1991imaging}
O.~Carnal, M.~Sigel, T.~Sleator, H.~Takuma, and J.~Mlynek.
\newblock Imaging and focusing of atoms by a fresnel zone plate.
\newblock {\em Physical Review Letters}, 67(23):3231--3234, 1991.

\bibitem{reisinger2009poisson}
T.~Reisinger, A.A. Patel, H.~Reingruber, K.~Fladischer, W.E. Ernst, G.~Bracco,
  H.I. Smith, and B.~Holst.
\newblock Poisson's spot with molecules.
\newblock {\em Physical Review A}, 79(5):053823, 2009.

\bibitem{sleator1992imaging}
T.~Sleator, T.~Pfau, V.~Balykin, and J.~Mlynek.
\newblock Imaging and focusing of an atomic beam with a large period standing
  light wave.
\newblock {\em Applied Physics B: Lasers and Optics}, 54(5):375--379, 1992.

\bibitem{pfau1997partial}
T.~Pfau and C.~Kurtsiefer.
\newblock Partial reconstruction of the motional wigner function of an ensemble
  of helium atoms.
\newblock {\em Journal of Modern Optics}, 44(11-12):2551--2564, 1997.

\bibitem{Herman80}
G.~T. Herman.
\newblock {\em Image Reconstruction from Projections: The Fundamentals of
  Computerized Tomography}.
\newblock Academic Press, New York, 1980.

\bibitem{Gaffney08062007}
K.~J. Gaffney and H.~N. Chapman.
\newblock Imaging atomic structure and dynamics with ultrafast x-ray
  scattering.
\newblock {\em Science}, 316(5830):1444--1448, 2007.

\end{thebibliography}

\end{document}